\documentclass[lettersize,journal]{IEEEtran}
\usepackage{amsmath,amsfonts}
\usepackage{algorithmic}
\usepackage{array}
\usepackage[caption=false,font=normalsize,labelfont=sf,textfont=sf]{subfig}
\usepackage{textcomp}
\usepackage{stfloats}
\usepackage{url}
\usepackage[T1]{fontenc}
\usepackage{verbatim}
\usepackage{cite}
\usepackage{color}
\usepackage{xcolor}
\usepackage{ulem}
\usepackage{lipsum}
\usepackage{makecell}
\usepackage{booktabs}
\usepackage{tabularx}
\usepackage{caption}
\usepackage{multirow}
\usepackage{graphicx}

\makeatletter
\let\MYcaption\@makecaption
\makeatother

\usepackage[hidelinks]{hyperref}
\makeatletter
\let\@makecaption\MYcaption
\makeatother

\def\BibTeX{{\rm B\kern-.05em{\sc i\kern-.025em b}\kern-.08em
    T\kern-.1667em\lower.7ex\hbox{E}\kern-.125emX}}
\usepackage{balance}
\usepackage{subfig}
\begin{document}
\title{Cooling Matters: Benchmarking Large Language Models and Vision-Language Models on Liquid-Cooled Versus Air-Cooled H100 GPU Systems}
\author{Imran Latif, Muhammad Ali Shafique, Hayat Ullah, Alex C. Newkirk, Xi Yu, Arslan Munir, \textit{Senior Member}, \textit{IEEE}
\thanks{Imran Latif is with Johnson Controls, Milwaukee, WI 53201, USA (e-mail: imran.latif@jci.com).}
\thanks{Muhammad Ali Shafique is with the Intelligent Systems, Computer Architecture, Analytics, and Security Laboratory (ISCAAS Lab), Department of Electrical and Computer Engineering, Kansas State University, Manhattan, KS 66506, USA (e-mail: alishafique@ksu.edu).}
\thanks{Hayat Ullah, and Arslan Munir are with the Intelligent Systems, Computer Architecture, Analytics, and Security Laboratory (ISCAAS Lab), Department of Electrical Engineering and Computer Science, Florida Atlantic University, Boca Raton, FL 33431, USA (e-mail: hullah2024@fau.edu, arslanm@fau.edu).}
\thanks{Alex C. Newkirk is with the Building Technology and Urban Systems, Lawrence Berkeley National Laboratory, Berkeley, CA 94720, USA (e-mail: acnewkirk@lbl.gov).}
\thanks{Xi Yu is with the Computing and Data Sciences Directorate, Brookhaven National Laboratory, Upton, NY 11973, USA (e-mail: xyu1@bnl.gov).}
}

\markboth{Journal of \LaTeX\ Class Files,~Vol.~X, No.~X, July~2025}%
{How to Use the IEEEtran \LaTeX \ Templates}

\maketitle

\begin{abstract}
The unprecedented growth in artificial intelligence (AI) workloads, recently dominated by large language models (LLMs) and vision-language models (VLMs), has intensified power and cooling demands in data centers. This study benchmarks LLMs and VLMs on two HGX nodes, each with 8× NVIDIA H100 graphics processing units (GPUs), using liquid and air cooling. Leveraging GPU Burn, Weights \& Biases, and IPMItool, we collect detailed thermal, power, and computation data. Results show that the liquid-cooled systems maintain GPU temperatures between 41–50°C, while the air-cooled counterparts fluctuate between 54–72°C under load. This thermal stability of liquid-cooled systems yields 17\% higher performance (54 TFLOPs/ GPU vs. 46 TFLOPs/GPU), performance-per-watt, reduced energy overhead, and greater system efficiency than the air-cooled counterparts. These findings underscore the energy and sustainability benefits of liquid cooling, offering a compelling path forward for hyperscale data centers seeking to optimize AI infrastructure. \href{https://github.com/iscaas/HPC-Performance-Benchmarking/tree/main/Liquid_vs_Air-cooled_H100_Benchmarking}{\textcolor{purple}{\texttt{https://github.com/iscaas/Cooling-Matters}}.}
\end{abstract}


\section{Introduction}
\label{sec:intro}
\IEEEPARstart{I}{n} recent years, the rapid development of large language models (LLMs) and vision-language models (VLMs) has dramatically transformed the landscape of artificial intelligence (AI), raising new challenges related to computational efficiency and thermal management in high-performance computing systems. As these models increase in complexity, the thermal output of the underlying hardware, particularly graphics processing units (GPUs) such as the NVIDIA H100, has become a critical factor determining energy performance. Effective cooling strategies are essential not only for maintaining optimal operating temperatures but also for ensuring consistent computational capabilities during compute intensive tasks, such as model training or image generation.\\
\begin{figure}[t]
  \centering
  \includegraphics[width=\linewidth]{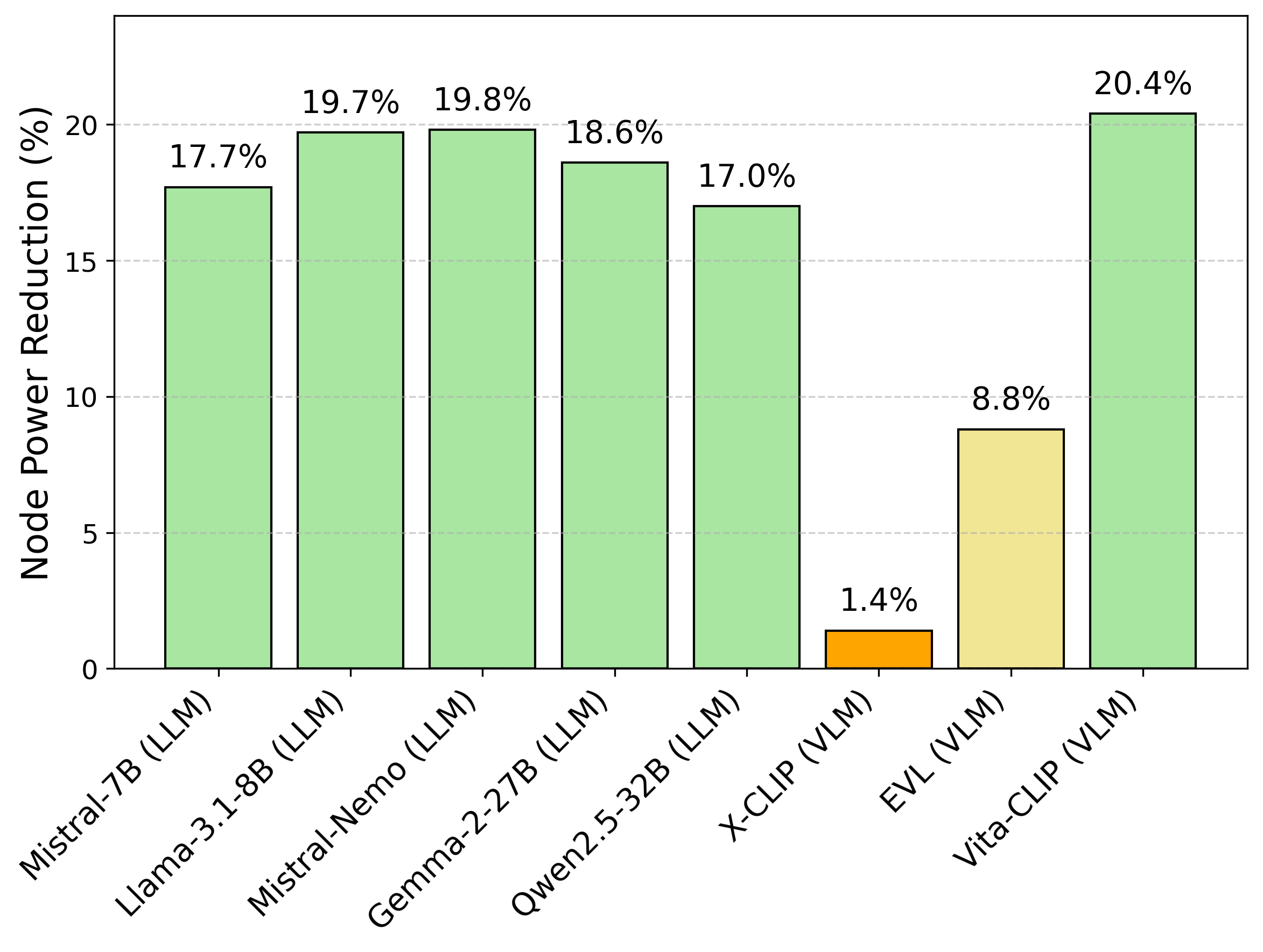}
  \caption{Node power reduction (\%) with liquid cooling over air cooling across LLM and VLM models.}
  \label{fig:power_reduction}
\end{figure}

\begin{figure*}[t]
  \centering
  \includegraphics[width=\linewidth]{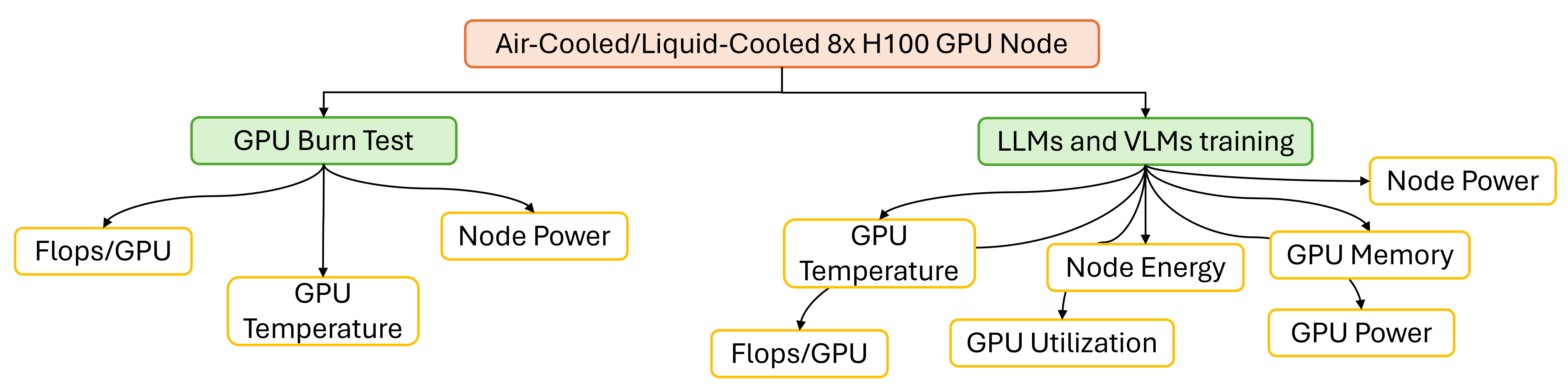}
  \caption{Benchmarking metrics for different cooling systems across various workloads.}
  \label{fig:benchmarks_on_workload}
\end{figure*}
\indent Today's rapid advancement of deep learning relies on algorithmic breakthroughs and high-performance computing infrastructure. As researchers uncover scaling laws linking model size,  total training data, and training computations~\cite{kaplan2020scalinglawsneurallanguage1}, the demand for specialized hardware capable of supporting large-scale workloads has surged. Training state-of-the-art deep learning models (i.e., LLMs and VLMs) with tens to hundreds of billions of parameters now routinely involves thousands of GPUs. Since late 2022, growing public interest and AI adoption have further accelerated this trend. In response, cloud providers have made substantial investments in AI infrastructure~\cite{rattner2024aiboom2}, anticipating sustained demand and future scaling needs. Hardware optimized for AI such as GPUs and tensor processing units (TPUs), offers high computational efficiency and has enabled transformative advances in scientific~\cite{schmidt2024gpus3} and industrial~\cite{nave2021artificial4} domains. However, the scaling of AI systems has led to substantial increases in power consumption and cumulative energy use, raising significant environmental concerns that require both monitoring and active mitigation strategies~\cite{kaack2022aligning5}.\\
\indent Growth in IT electricity demand also requires cooling infrastructure to keep pace. Traditional air-cooled systems have struggled to keep up with the escalating thermal demands imposed by modern GPUs~\cite{narayanan2024investigation}. Liquid cooling, including direct-to-chip and immersion cooling techniques, has been proven to be more effective in handling high power densities characteristic of GPU systems like the NVIDIA H100, demonstrating significant improvements in thermal performance and energy efficiency~\cite{ohenhen2024sustainable, zhou2024immersion, naduvilakath2024numerical}. For instance, research indicates that liquid-cooled systems can successfully lower the GPUs temperature by reducing thermal resistance, thus enhancing overall system performance while minimizing energy consumption~\cite{ohenhen2024sustainable}. The implementation of innovative methodologies, such as fuzzy control algorithms for liquid cooling systems, further demonstrates the evolving strategies proposed for optimizing thermal management in high-performance computing environments~\cite{wang2024research}.\\
\indent Moreover, the effective benchmarking of AI models necessitates the evaluation of thermal performance of the underlying hardware. Evaluations have shown that the efficiency of LLMs and VLMs is significantly impacted by thermal dynamics within GPU systems~\cite{naduvilakath2024numerical, shao2020evaluation}. As models continue to grow in complexity and size, ensuring adequate cooling becomes crucial not only for the hardware's longevity but also for the accuracy and efficiency of model inference. The integration of various cooling techniques, including advanced liquid cooling solutions, can potentially enhance the performance of computational tasks by maintaining optimal operating temperatures~\cite{zhou2024immersion}.\\
\indent In this work, we present a detailed benchmarking study that evaluates the impact of cooling strategy such as liquid versus air, on the performance and power efficiency of AI workloads using 8$\times$ NVIDIA H100 GPU nodes. Our analysis includes both synthetic stress testing with \textit{GPU-Burn} and real-world model training across five LLMs: Mistral-7B-v0.3~\cite{jiang2023mistral7b}, LLaMA-3.1-8B~\cite{grattafiori2024llama}, Mistral-NeMo-Base-2407~\cite{mistralnemo2024}, Gemma-2-27B~\cite{gemmateam2024gemma2improvingopen}, and Qwen2.5-32B~\cite{yang2024qwen2}, as well as three VLMs: X-CLIP~\cite{ni2022expanding}, EVL~\cite{lin2022frozen}, and Vita-CLIP~\cite{wasim2023vita}. We measure a comprehensive set of metrics, including GPU power draw, temperature, utilization, training duration, energy consumption, achieved FLOPs, and GPU-level and node-level power consumption. By analyzing thermal, computational, and power behavior across cooling architectures, our results provide insight into how cooling infrastructure affects performance-per-watt under realistic AI workloads as shown in Figure~\ref{fig:power_reduction}.\\
\textbf{Contributions:} The main contributions of this work are as follows:
\begin{enumerate}
  \item To the best of our knowledge, we present one of the first empirical comparisons of LLM and VLM training workloads on liquid-cooled and air-cooled nodes having 8$\times$ NVIDIA H100 GPUs) under same GPU hardware configurations.
  
  \item We introduce a detailed benchmarking methodology combining synthetic stress tests, instruction fine-tuning, VLMs training, and real-time GPU and node-level power profiling using tools such as \textit{ipmitool}, \textit{CodeCarbon}, and \textit{Weights \& Biases}.
  
  \item We thoroughly quantify the thermal and power efficiency benefits of liquid cooling, showing up to 17\% higher TFLOPs per GPU and over 1 kW lower node-level power draw across diverse training workloads.
  
  
  \item We provide detailed insights into workload-level power and energy characteristics for scalable AI infrastructure, offering a reproducible framework for future energy-aware benchmarking efforts.
\end{enumerate}

\noindent\textbf{Significant Results:} The most significant results of our benchmarking study are summarized as follows:
\begin{enumerate}
    \item Liquid-cooled systems maintained GPUs temperature between 41–50\textdegree C under peak load as compared to 54–72\textdegree C in air-cooled systems.
  \item During GPU Burn, the liquid-cooled node achieved an average of 54 TFLOPs/GPU, 17\% higher than the 46 TFLOPs/GPU observed on the air-cooled node.
  \item During LLM fine-tuning and VLM (specifically ViTA-CLIP) training, node-level power consumption on the liquid-cooled system is consistently lower by 1–1.5 kW, while maintaining equal or improved training duration.
  \item All LLM and VLM models demonstrated higher performance-per-watt on the liquid-cooled system, as compared to air-cooled system.
  \item All models maintained equal or improved training durations on liquid-cooled systems and exhibited higher throughput compared to their air-cooled counterparts. 
\end{enumerate}

The rest of this paper is structured as follows. In Section~\ref{sec:background}, we provide background on the evolution of AI infrastructure and the motivation for evaluating cooling strategies in high-performance training environments. Section~\ref{sec:methodology} outlines the methodology used to benchmark system performance, power, and thermal behavior. Section~\ref{sec:experiment} describes the experimental setup, including hardware specifications, software configuration, and workload design. In Section~\ref{sec:results}, we present the measured power, temperature, and performance metrics across both cooling configurations. Section~\ref{sec:discussion} analyzes the findings in the context of system design and energy efficiency. Finally, Section~\ref{sec:conclusion} concludes this paper and discusses broader implications for sustainable AI infrastructure and directions for future work.

\section{Background and Motivation}
\label{sec:background}
The performance and efficiency of advanced deep learning models, particularly large language models (LLMs) and vision-language models (VLMs) are deeply influenced by the thermal and energy behavior of the hardware on which they run. As these models continue to grow in complexity and scale, effective thermal management becomes increasingly critical to avoid bottlenecks associated with heat dissipation and power consumption. This study is motivated by the need to understand how cooling infrastructure, specifically liquid versus air cooling, affects energy efficiency, computational throughput, and sustainability when training and deploying large-scale models.\\
\indent Cooling systems can account for 30\% to 50\% of the total energy use in data centers, with air cooling still being the dominant technique~\cite{kim2024data,tang2023experimental}. However, the rising thermal output of modern GPUs such as the NVIDIA H100 has exposed the limitations of traditional air-cooled systems in high-density environments. Given the increasing energy demands of training and inference for LLMs and VLMs, understanding and improving cooling efficiency is of paramount importance~\cite{tang2023experimental}. Liquid cooling, especially direct-to-chip and immersion techniques has emerged as a more thermally efficient alternative, capable of reducing energy consumption by as much as 50\% in some cases~\cite{haghshenas2023enough}. These reductions are not only energy-saving but also help prevent thermal throttling, which can degrade GPU performance under sustained load~\cite{icae2021freecooling}.\\
\indent Thermal benchmarking also necessitates a careful analysis of trade-offs between cooling efficiency, operational uptime, and resource utilization. Gupta et al.~\cite{gupta2021energy} emphasized that intelligent cooling strategies can influence both energy and power efficiency, optimizing resource allocation at the data center level. In this context, our benchmarking efforts aim to contribute empirical insights by comparing liquid- and air-cooled  H100 nodes (having 8$\times$ H100 GPUs) using real-world LLM and VLM training workloads. Energy usage effectiveness (PUE) can also serve as a key performance metric in assessing the relative efficiency of the two cooling approaches~\cite{kim2024data}.\\
\indent The shift toward intelligent and sustainable thermal management is also evident in the use of predictive and adaptive algorithms to control cooling systems. Recent works explore how machine learning, fuzzy control, and evolutionary algorithms can optimize cooling in real time without compromising compute performance~\cite{wang2022toward, athavale2021}. These methods are particularly relevant as operational costs and the carbon footprint of large-scale AI infrastructure become major concerns for both industry and academia~\cite{ebirim2024optimizing}.\\
\indent The challenges of thermal regulation become even more critical in the context of high-density workloads of LLMs and VLMs. Conventional air-cooling approaches often fall short in managing the intense heat generated under prolonged GPU utilization. Recent studies have demonstrated that advanced cooling architectures, such as phase-change systems and immersion cooling offer superior thermal regulation, particularly in environments with extreme external temperatures~\cite{zhao2023simulation, jia2024simulation}. These technologies not only support performance stability but also reduce the need for costly cooling redundancy, making them attractive for sustainable deployment.\\
\indent In summary, recent literature highlights a growing consensus: cooling infrastructure plays a central role in shaping the performance, energy consumption, and sustainability of AI workloads. Our study contributes to this domain by empirically comparing liquid- and air-cooled H100 GPU systems under representative LLM and VLM workloads. Through this benchmarking, we aim to offer actionable insights into how thermal management can influence both the efficiency and reliability of modern AI systems.

\section{Methodology}
\label{sec:methodology}
This section outlines the benchmarking methodology used to evaluate air-cooled and liquid-cooled system performance, power consumption, and thermal behavior. We detail the training workloads, synthetic stress test, measurement tools, and profiling strategies used to capture GPU-level and node-level metrics. The approach is designed to isolate the impact of cooling systems on our profiling mechanism across all tests.\\
\indent To assess maximum thermal load and theoretical peak compute performance, we used \textit{GPU Burn}, a high-intensity CUDA stress test. The utility was run simultaneously on all 8 GPUs across both nodes, generating consistent and repeatable full-load conditions. \textit{GPU Burn} provided automatic reporting of TFLOPS per GPU, which we used to evaluate compute throughput. During each test, we recorded individual GPU temperatures, power draw, and utilization, as well as total node power using \textit{ipmitool}. This allowed us to establish upper bounds for system performance and energy demand under sustained maximum load.\\
\indent The GPU Burn test served as a reference point against which real-world LLM and VLM training workloads were compared, highlighting the difference between synthetic and practical energy-performance profiles. To evaluate the thermal and computational behavior of both LLM and VLM training workloads, we captured a comprehensive set of node-level and GPU-level metrics throughout each run as shown in Figure~\ref{fig:benchmarks_on_workload}. The measurements included:
\begin{itemize}
  \item \textbf{GPU Temperature}: Per-GPU temperature readings are collected continuously to monitor thermal stability across both liquid- and air-cooled systems.
  \item \textbf{Memory Usage}: Peak and average memory utilization per GPU is recorded to assess memory load distribution during training of various models.
  \item \textbf{GPU Utilization}: Real-time utilization metrics were used to ensure full workload saturation and to compare operational efficiency across nodes.
  \item \textbf{Individual GPU Power Draw}: Power consumption is tracked for each GPU, enabling localized power efficiency analysis.
  \item \textbf{Average GPU Power}: Aggregate power usage across all GPUs in a node is computed for a high-level comparison of cooling impact.
  \item \textbf{Average Node Power}: Total node power is sampled via IPMI at 20-second intervals, providing an external view of system power and energy consumption.
  \item \textbf{FLOPS per GPU}: Computational throughput (in TFLOPs and GFLOPs) is measured using DeepSpeed FLOP Profiler and THOP for training workloads, and GPU Burn’s internal profiler for synthetic tests.
\end{itemize}
\indent \indent This multi-dimensional profiling allowed for a detailed comparison of performance, energy consumption, and thermal behavior across cooling configurations and model types.

\section{Experimental Setup and Equipment}
\label{sec:experiment}
This section outlines the hardware infrastructure and software frameworks used for benchmarking. It includes the details of liquid-cooled and air-cooled nodes, followed by the software stack used to monitor, and profile metrics across all experiments.
\subsection{Computational Hardware}
Our experiments were conducted on two high-end GPU nodes with identical GPU configurations but different cooling systems, that is, liquid-cooled and air-cooled setups.
\subsubsection{Liquid-Cooled Node}: This system contains 8$\times$ NVIDIA H100 80GB HBM3 GPUs, supported by an AMD EPYC 9474F 48-core processor with 192 threads. This node includes a liquid-cooled system to maintain optimal thermal regulation.
\subsubsection{Air-Cooled Node}: This node was also featured with 8× NVIDIA H100 80GB HBM3 GPUs, but it uses an AMD EPYC 9534 64-core processor with 256 threads. It provided an identical 1.5 TB RAM and operated on air-cooled system.
\subsubsection{Cooling Configuration}:
The liquid-cooled node uses direct-to-chip (D2C) cooling for both CPU and GPU, with CoolIT Systems OAT PG-25 coolant circulated through a closed-loop system. The circuit volume is 16.3 liters (4.3 US gallons), and the system operates at a maximum return pressure of 50 PSI, with a secondary loop pressure relief cap at 86 PSI. Peripheral components such as memory and storage remain air-cooled in both configurations. The air-cooled node contains 8 fans, while the liquid-cooled node uses only 4 fans, as fewer components require active airflow. This reduction contributes to lower power consumption and reduced acoustic noise.
\subsubsection{Datacenter Cooling Infrastructure}:
The datacenter utilizes a shared chilled water loop maintained at 20°C, produced using conventional chillers with cooling towers. There is no dedicated chiller exclusively for the DLC nodes. The air-cooled system uses a rear-door heat exchanger, while the DLC node connects directly to the chilled water loop via cold plates. All tests were conducted at ambient room temperatures of 21–22°C.
\subsection{Software Configuration}
We used a variety of tools to monitor key metrics at the GPU level and the node level. It includes GPU memory, power, temperature, utilization, and power consumption at the GPU level and node level. Each tool provides specific measurements essential for comparing air-cooled and liquid-cooled systems.
\subsubsection{IPMITools}
It is a command-line utility to access hardware-level metrics using the Intelligent Platform Management Interface (IPMI) \cite{ipmitool}. In our benchmarking across all workloads, it was used to log node-level power consumption at the interval of 20 seconds during GPU Burn test and LLM/VLM training. Although GPU-specific metrics were measured separately, ipmitool provides a node-level power consumption for the analysis of performance-per-watt across liquid-cooled and air-cooled configurations.
\subsubsection{Weight \& Biases}
It is the machine learning platform that provides tools for training, tracking, and visualizing experiments data, model, and metrics \cite{wandb}. In our work, it was used to capture GPU hardware metrics such as memory, temperature, utilization, and power consumption during LLM/VLM training. These measurements helped in comparing the performance of nodes with different cooling configurations.
\subsubsection{CodeCarbon}
It is an open-source tool for estimating carbon emission with average power and energy consumption for CPU, GPU and RAM. In our experiments, it was used to track average GPU power and energy during LLM/VLM training. Unlike node-level power measurement, \textit{CodeCarbon} \cite{benoit_courty_2024_11171501} provides GPU-specific average power and energy metrics in Watt and Watt-hours (Wh) respectively, which were used for comparing of air-cooled and liquid-cooled nodes for various models.
\subsubsection{FLOPS Profiler}
To measure the FLOPS for LLMs and VLMs, we used the \textit{DeepSpeed FLOP Profiler} and \textit{THOP}. The DeepSpeed profiler provided runtime FLOP estimates during the training. This tool enabled us to estimate FLOPS per GPU during LLM/VLM training for air-cooled and liquid-cooled node. While GPU Burn test scripts provides the FLOPS per GPU automatically. The results collected from profilers discussed above, allowed us to compare the compute efficiency and the performance-per-watt across various cooling configurations.
\subsection{Workload Configuration}
\subsubsection{GPU Burn Test Setup}
To compare the performance difference of air-cooled and liquid-cooled nodes under maximum workload, we employed the GPU Burn~\cite{timonen2024gpuburn} utility. We operated this test by enabling double-precision computation. The test was run for two hours on all 8 GPUs on each node using the following configuration:
\begin{center}
\textit{gpu-burn -d 7200}
\end{center}

\begin{figure}[t]
  \centering
  \includegraphics[width=\linewidth]{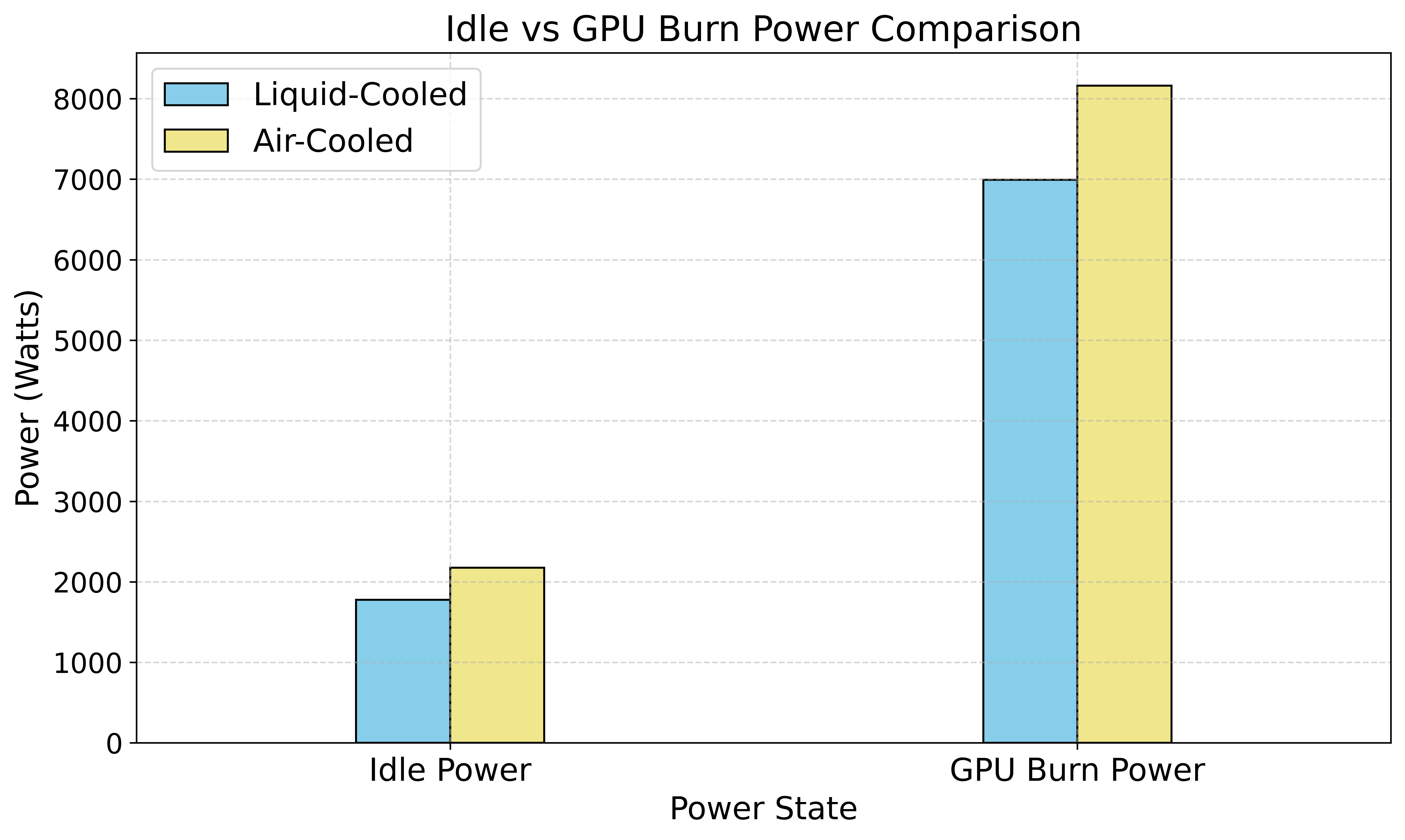}
  \caption{Power comparison for GPU Burn Test on air-cooled and liquid-cooled nodes.}
  \label{fig:gpu_burn_power}
\end{figure}
\begin{figure}[t]
  \centering
  \includegraphics[width=\linewidth]{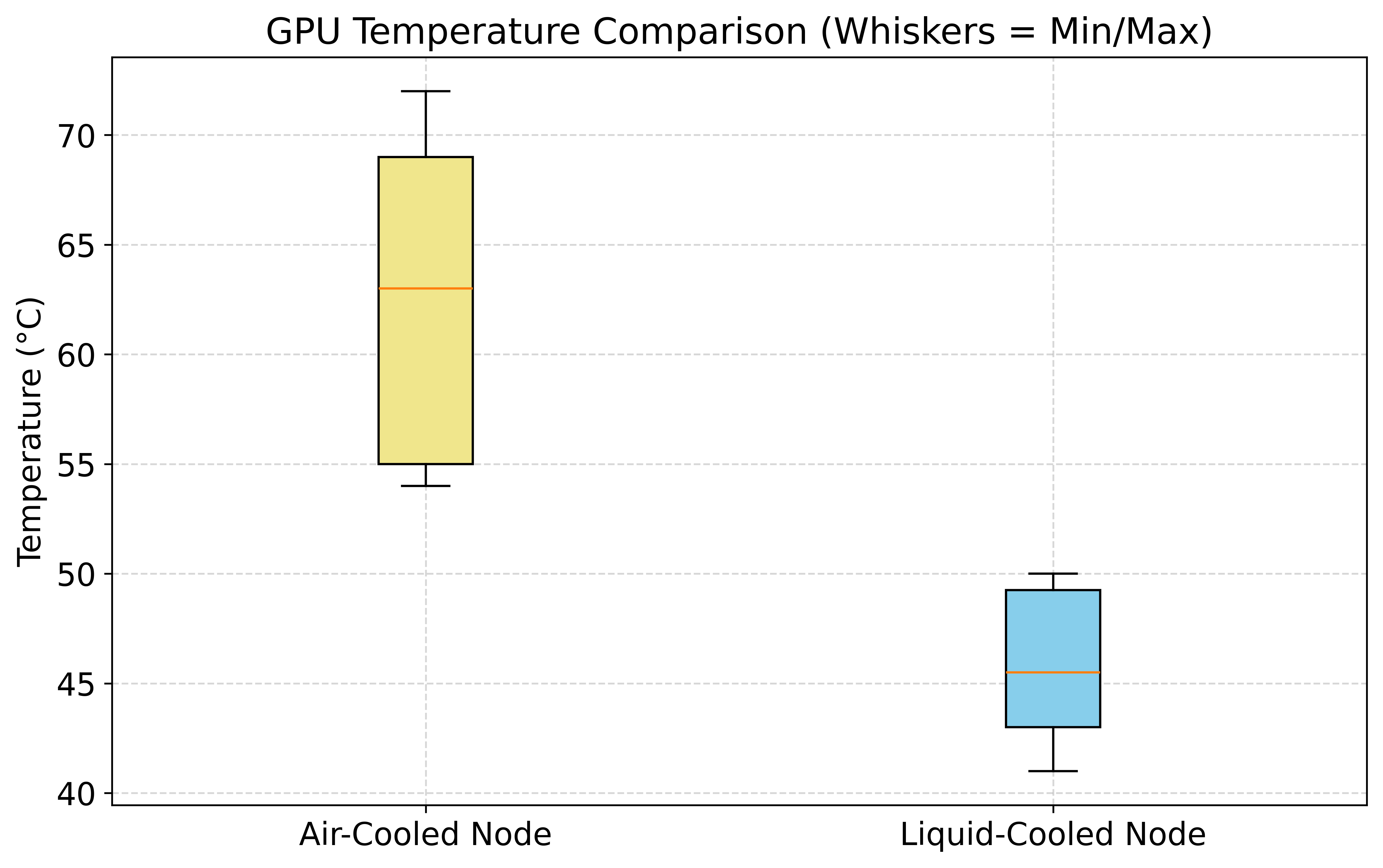}
  \caption{Temperature comparison for GPU Burn Test on air-cooled and liquid-cooled nodes.}
  \label{fig:gpu_burn_temp}
\end{figure}

\begin{figure}[t]
  \centering
  \includegraphics[width=\linewidth]{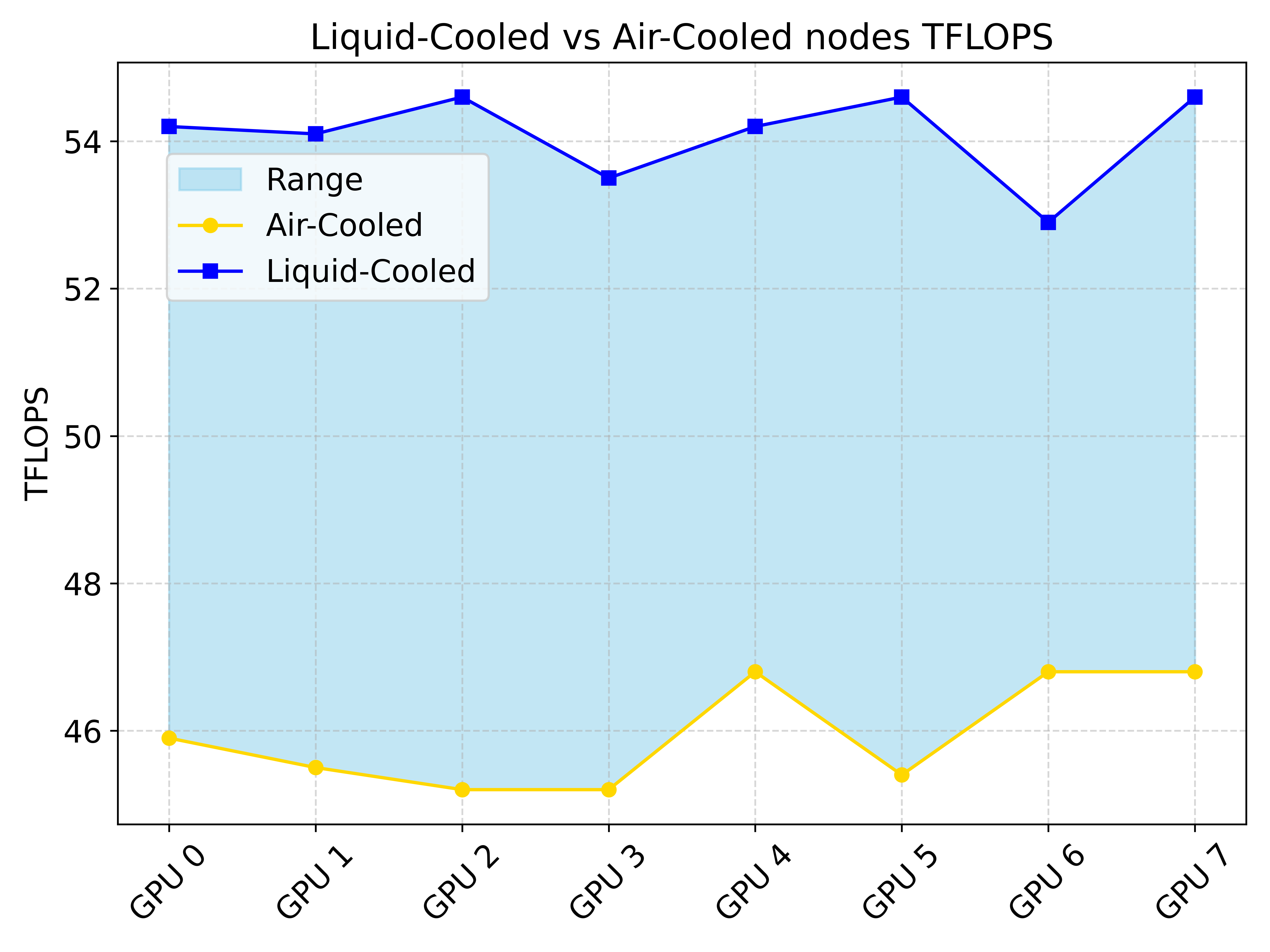}
  \caption{TFLOPS difference for GPU Burn test on air-cooled and liquid-cooled nodes.}
  \label{fig:gpu_burn_flops}
\end{figure}

\indent \indent In above command, \textit{-d 7200} sets the duration of GPU Burn test to 7200 seconds (2 hours). This configuration was selected to test each node under continuous high-intensity workloads. During the test, we measured each GPU temperature and FLOPS by logging GPU Burn test results. Total node power was also monitored during this test using \textit{ipmitool}. These results served as an important reference point for evaluating the efficiency and thermal resilience of each cooling node under maximum workload.
\subsubsection{LLM Training Setup}
All pre-trained large language models were finetuned using the Alpaca~\cite{alpaca} dataset in a supervised manner. Distributed data parallel (DDP) technique was used across 8$\times$ NVIDIA H100 air-cooled and liquid-cooled nodes. Due to limited computational resource, parameter-efficient finetuning (PEFT) such as QLoRA was employed, with a low-rank value $r = 128$. Each model was trained for one epoch with a micro batch size of 4. With DDP running on 8$\times$ GPU, the effective batch size was 32. This training setup was consistent, with each LLM finetuning, allowing consistent comparisons across air-cooled and liquid-cooled nodes.
\subsubsection{VLM Training Setup}
To systematically evaluate and compare the performance capabilities of liquid-cooled and air-cooled computing nodes, we conducted comprehensive training experiments using three VLM models: X-CLIP~\cite{ni2022expanding}, EVL~\cite{lin2022frozen}, and ViTA-CLIP~\cite{wasim2023vita} on UCF101~ \cite{soomro2012ucf101} dataset. All models are trained with the same hyper-parameters on both nodes to ensure a fair comparison. Specifically, each model is trained in Distributed Data-Parallel (DDP) settings across 8$\times$ NVIDIA H100 liquid-cooled and air-cooled nodes.\\

\indent Each model is trained for 50 epochs with a global batch size of 32 and a mini-batch size of 4 (i.e., 4 samples per GPU). Additionally, we set the video sequence length to 16 frames for each model. These hyperparameters are consistently maintained across both liquid-cooled and air-cooled nodes to ensure a fair and accurate performance comparison. This experimental setup enables a detailed assessment of the training efficiency, computational performance, and resource utilization differences between the liquid-cooled and air-cooled nodes.

\section{Results}
\label{sec:results}
This section presents performance and power measurements collected during LLM/VLM training and GPU Burn test. We used these results to evaluate both cooling systems under various workloads.

\begin{table*}[t]
  \centering
  \caption{Training metrics across various LLM models for air/liquid-cooled nodes.}
  \label{tab:llm_training_metrics}
  \begin{tabular}{cccccccc}
    \toprule
    \multirow{2}{*}{Model} & \multirow{2}{*}{Node} & Trainable (\%) & Duration & Avg. 8$\times$ GPU & Avg. 8$\times$ GPU & Avg. Node Power & \multirow{2}{*}{TFLOPS/GPU} \\
    & & / Params & (sec) & Utilization (\%) & Power (W) & (W) & \\
    \midrule
    Mistral-7B-v0.3~\cite{jiang2023mistral7b}& Liquid-Cooled & 4.42\% / 7B & 2702.03 & 95.8 & 3503 & 5051 & 35.19 \\
               & Air-Cooled        & 4.42\% / 7B & 2721.13 & 95.3 & 3658 & 6135 & 35.09 \\
    \midrule
    Llama-3.1-8B~\cite{grattafiori2024llama} & Liquid-Cooled & 4.01\% / 8B & 2466.19 & 94.9 & 3505 & 4975 & 35.12 \\
                 & Air-Cooled        & 4.01\% / 8B & 2479.83 & 95.2 & 3653 & 6193 & 34.76 \\
    \midrule
    Mistral-Nemo- & Liquid-Cooled & 3.59\% / 12B & 3827.89 & 95.8 & 3556 & 5052 & 35.71 \\
                     Base-2407~\cite{mistralnemo2024}& Air-Cooled        & 3.59\% / 12B & 3843.79 & 95.6 & 3717 & 6296 & 35.52 \\
    \midrule
    Gemma-2-27B~\cite{gemmateam2024gemma2improvingopen} & Liquid-Cooled & 3.24\% / 27B & 8393.51 & 92.7 & 3742 & 5333 & 39.17 \\
                & Air-Cooled        & 3.24\% / 27B & 8418.65 & 93.1 & 3912 & 6550 & 39.15 \\
    \midrule
    Qwen2.5-32B~\cite{yang2024qwen2} & Liquid-Cooled & 3.17\% / 32B & 9738.14 & 93.4 & 3671 & 5344 & 37.42 \\
                & Air-Cooled        & 3.17\% / 32B & 9783.15 & 92.9 & 3844 & 6440 & 37.40 \\
  \bottomrule
\end{tabular}
\end{table*}

\begin{figure*}[t]
  \centering
  \subfloat[\label{fig:Qwen_gpu_power_sm}]{%
    \includegraphics[width=0.32\textwidth]{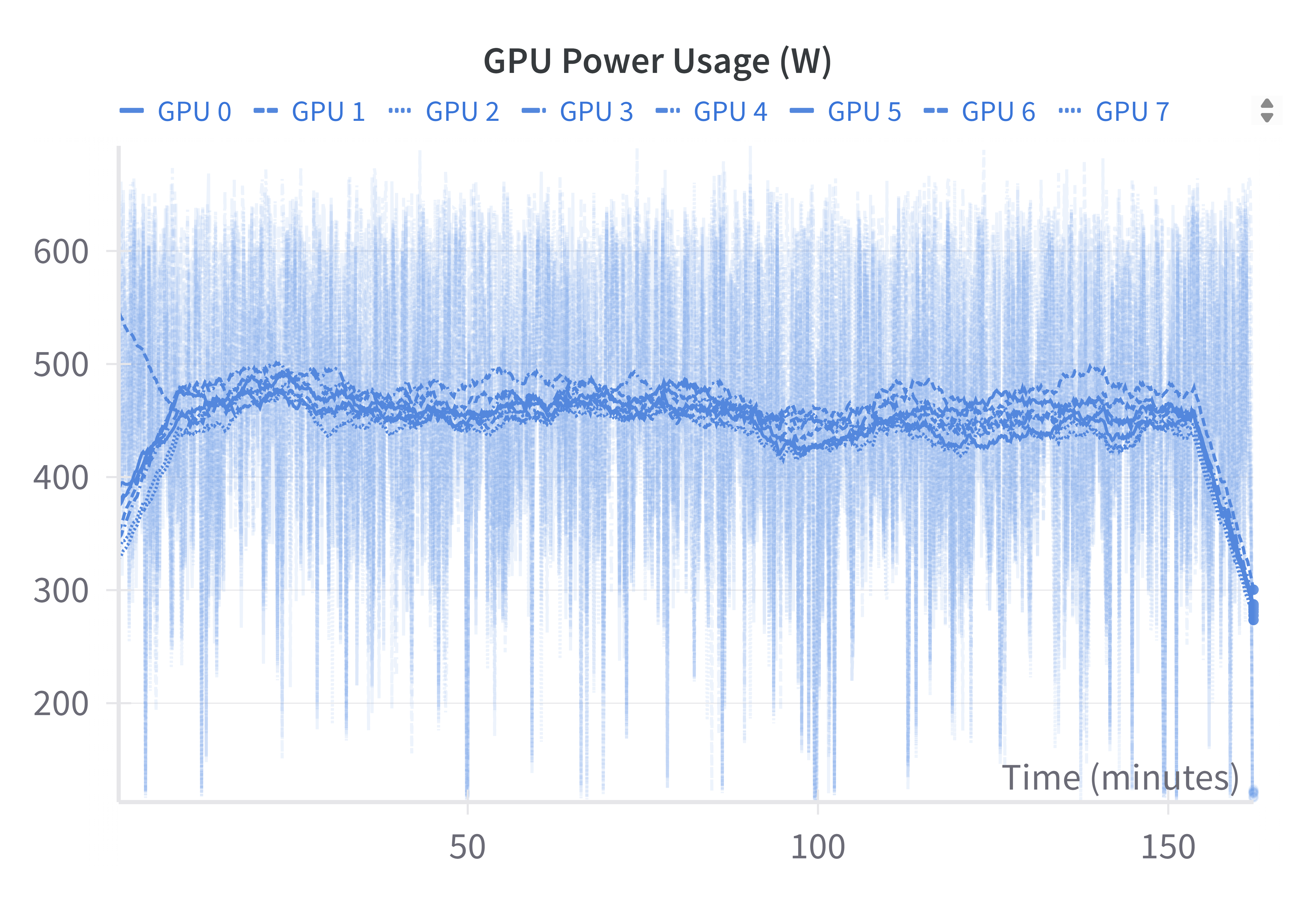}
  }
  \hfill
  \subfloat[\label{fig:Qwen_gpu_temp_sm}]{%
    \includegraphics[width=0.32\textwidth]{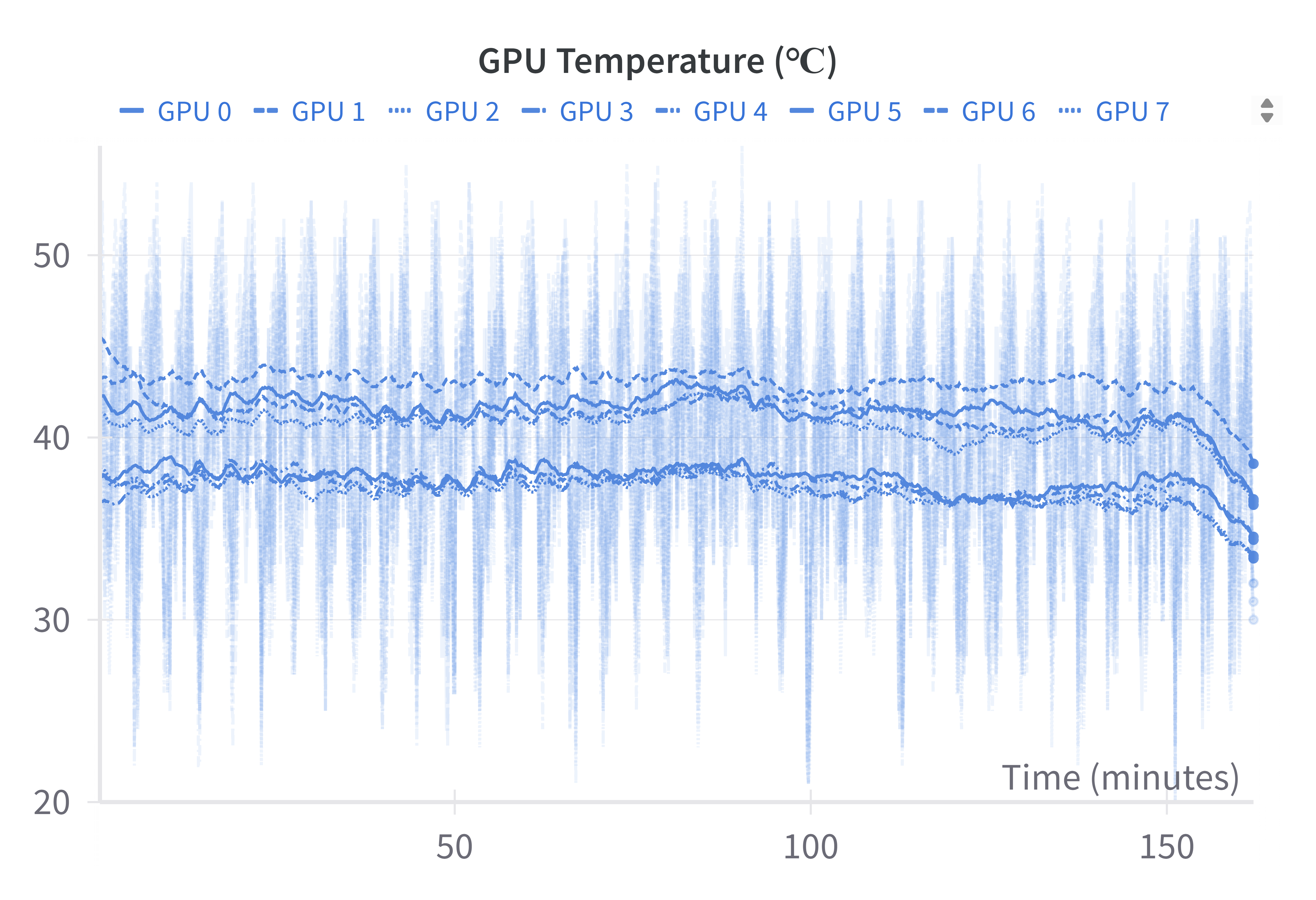}
  }
  \hfill
  \subfloat[\label{fig:Qwen_gpu_util_sm}]{%
    \includegraphics[width=0.32\textwidth]{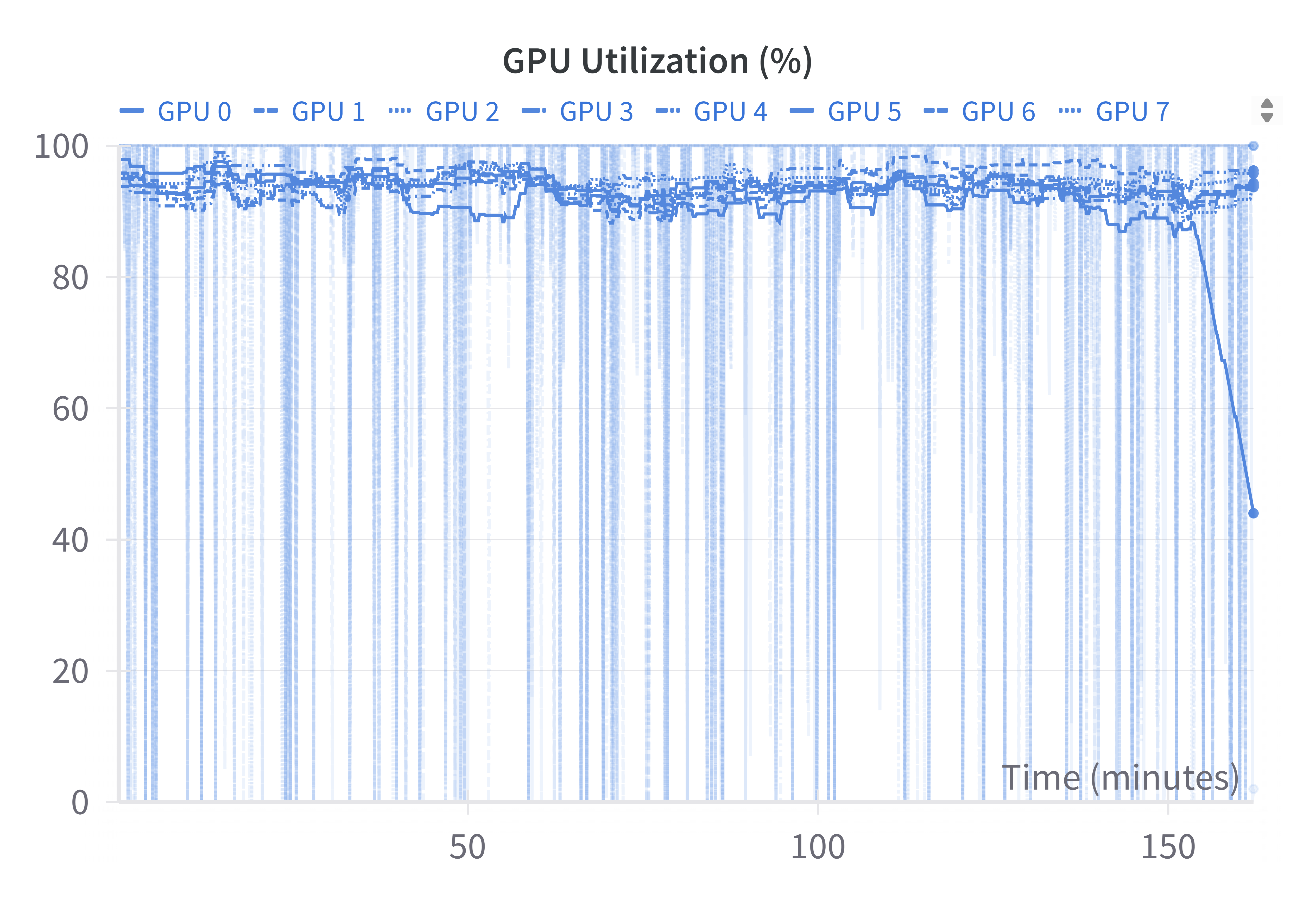}
  }

  \vspace{0.5em}

  \subfloat[\label{fig:Qwen_gpu_power_bnl}]{%
    \includegraphics[width=0.32\textwidth]{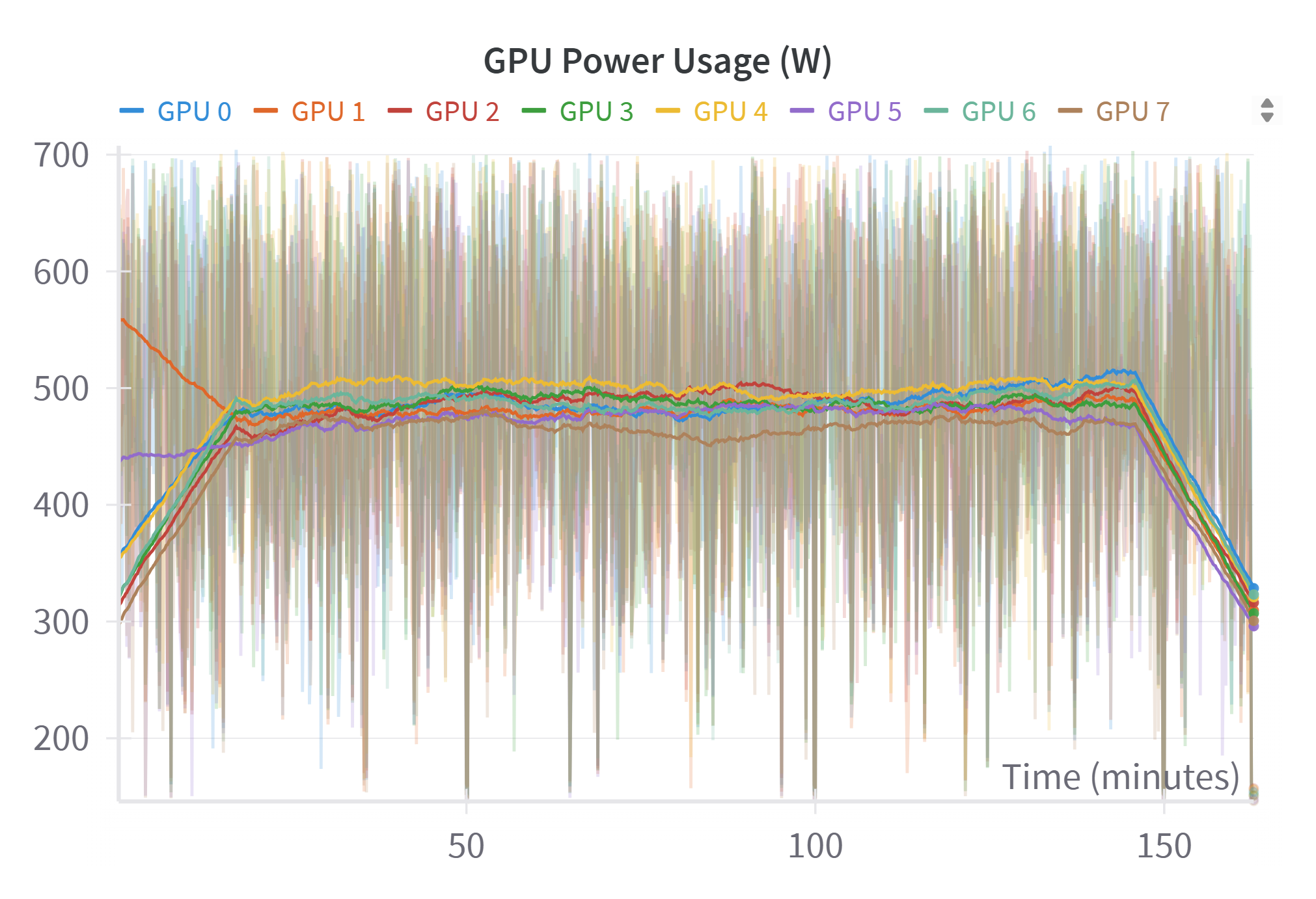}
  }
  \hfill
  \subfloat[\label{fig:Qwen_gpu_temp_bnl}]{%
    \includegraphics[width=0.32\textwidth]{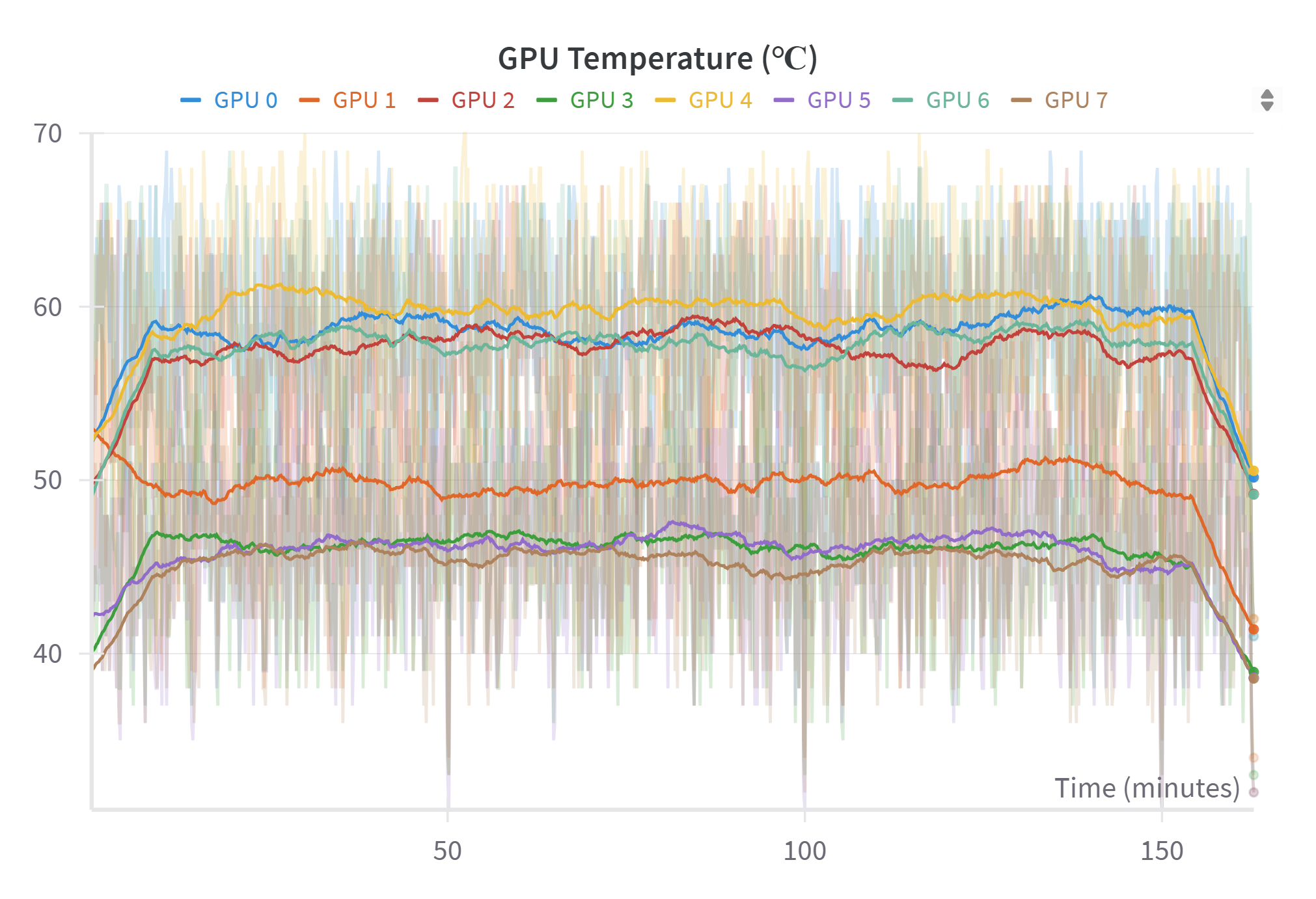}
  }
  \hfill
  \subfloat[\label{fig:Qwen_gpu_util_bnl}]{%
    \includegraphics[width=0.32\textwidth]{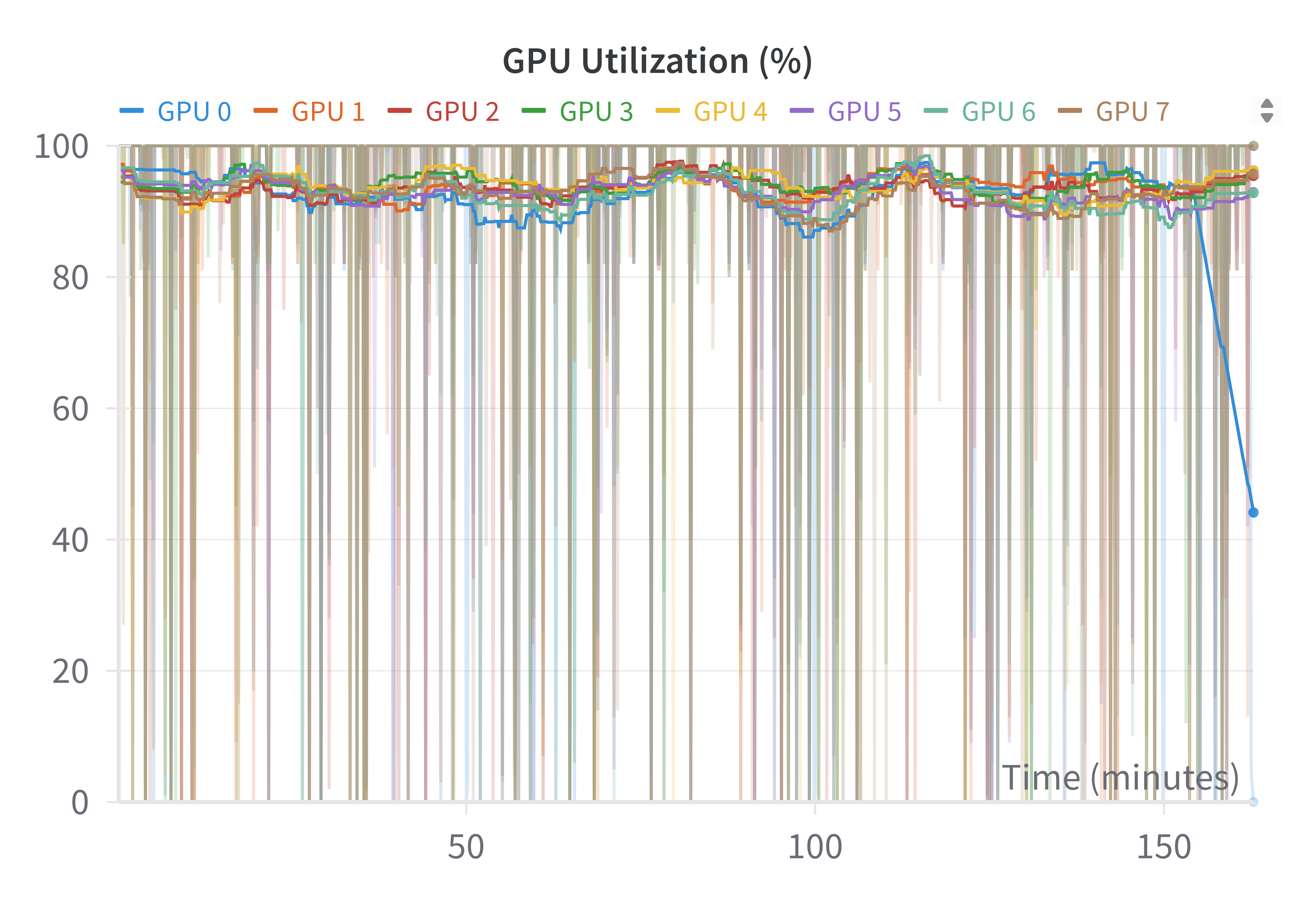}
  }

  \caption{GPU metrics comparison for liquid-cooled (a–c) and air-cooled (d–f) nodes for Qwen2.5-32B~\cite{yang2024qwen2}.}
  \label{fig:six_figures_qwen}
\end{figure*}

\subsection{GPU Burn Test}
To evaluate peak throughput and thermal performance of both cooling systems, we conducted GPU Burn test on both air-cooled and liquid-cooled nodes. We observed that during the idle state, the liquid-cooled node consumed 1.78 kW on average, as compared to air-cooled node which consumed 2.17 kW. For full load, power consumption for liquid-node was 6.99 kW on average while air-cooled node power usage was 8.16 kW. These results showed that as the load increased, the power difference between liquid-cooled and air-cooled node increased significantly as shown in Figure~\ref{fig:gpu_burn_power}.\\
\indent Similar trend was observed for GPU temperatures during peak workload. The GPUs in air-cooled node reached up to 72°C, whereas GPUs of liquid-cooled node remained between 41–50°C. In addition, temperature of air-cooled GPUs range from 54°C to 72°C, larger than the temperature range of liquid-cooled GPUs as depicted in Figure~\ref{fig:gpu_burn_temp}. This temperature peak and range difference across both cooling configurations translated into higher throughput on liquid-cooled node, with per-GPU FLOPS ranging from 52.9–54.6 TFLOPS as compared to 45.2–46.8 TFLOPs for air-cooled node, exhibiting a clear thermal advantage for the liquid-cooled system as illustrated in Figure~\ref{fig:gpu_burn_flops}.
\\
\indent These results underscore the performance advantage of the liquid cooling system, particularly for high-intensity workloads. Not only liquid-cooled node consumed less power at peak workload, but it also delivered approximately 17\% more compute per GPU while maintaining significantly lower temperatures.

\begin{table*}[t]
  \centering
  \caption{Training metrics across various VLM models for liquid-cooled and air-cooled nodes.}
  \label{tab:VLM_results}
  \begin{tabular}{cccccccc}
    \toprule
    Model & Node & \makecell{Trainable\% \\ / Params} & \makecell{Duration\\(sec)} & \makecell{Avg. 8$\times$ GPU \\Utilization (\%)} & \makecell{Avg. 8$\times$ GPU\\Power (W)} & \makecell{Avg. Node Power\\(W)} & GFLOPS/GPU \\
    \midrule
    \multirow{2}{*}{\rotatebox[origin=c]{0}{X-CLIP \cite{ni2022expanding}}} & Liquid-Cooled & 100\% / 97.95M & 1503 & 58.6 & 3268 & 5890 & 377.13 \\
               & Air-Cooled        &  100\% / 97.95M & 1549 & 58.1 & 3383 & 5975 & 376.92 \\
    \midrule
   \multirow{2}{*}{\rotatebox[origin=c]{0}{EVL \cite{lin2022frozen}}} & Liquid-Cooled & 100\% / 28.43M & 14530 & 60.0 & 2320 & 4053 & 119.87 \\
                 & Air-Cooled        & 100\% / 28.43M  & 15296 & 59.3 & 2351 & 4444 & 119.41 \\
    \midrule
    \multirow{2}{*}{\rotatebox[origin=c]{0}{Vita-CLIP \cite{wasim2023vita}}} & Liquid-Cooled & 100\% / 146.24M & 91654 & 97.4 & 4295 & 5867 & 1064.06 \\
    & Air-Cooled        &  100\% / 146.24M & 91941 & 97.8 & 4529 & 7375 & 1063.81 \\
  \bottomrule
\end{tabular}
\end{table*}

\begin{figure*}[t]
  \centering
  \subfloat[\label{fig:vclip_gpu_power_sm}]{%
    \includegraphics[width=0.32\textwidth]{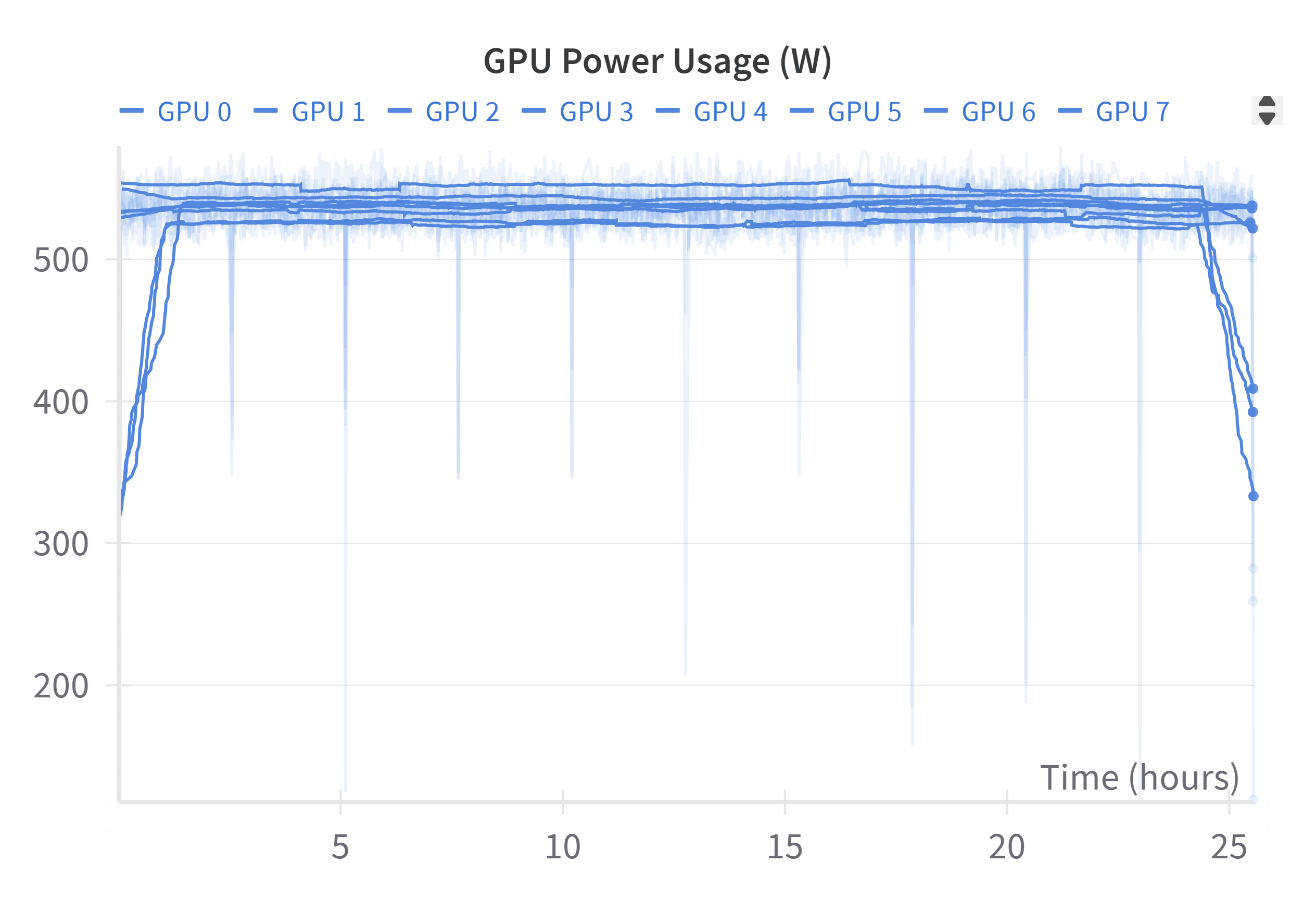}
  }
  \hfill
  \subfloat[\label{fig:vclip_gpu_temp_sm}]{%
    \includegraphics[width=0.32\textwidth]{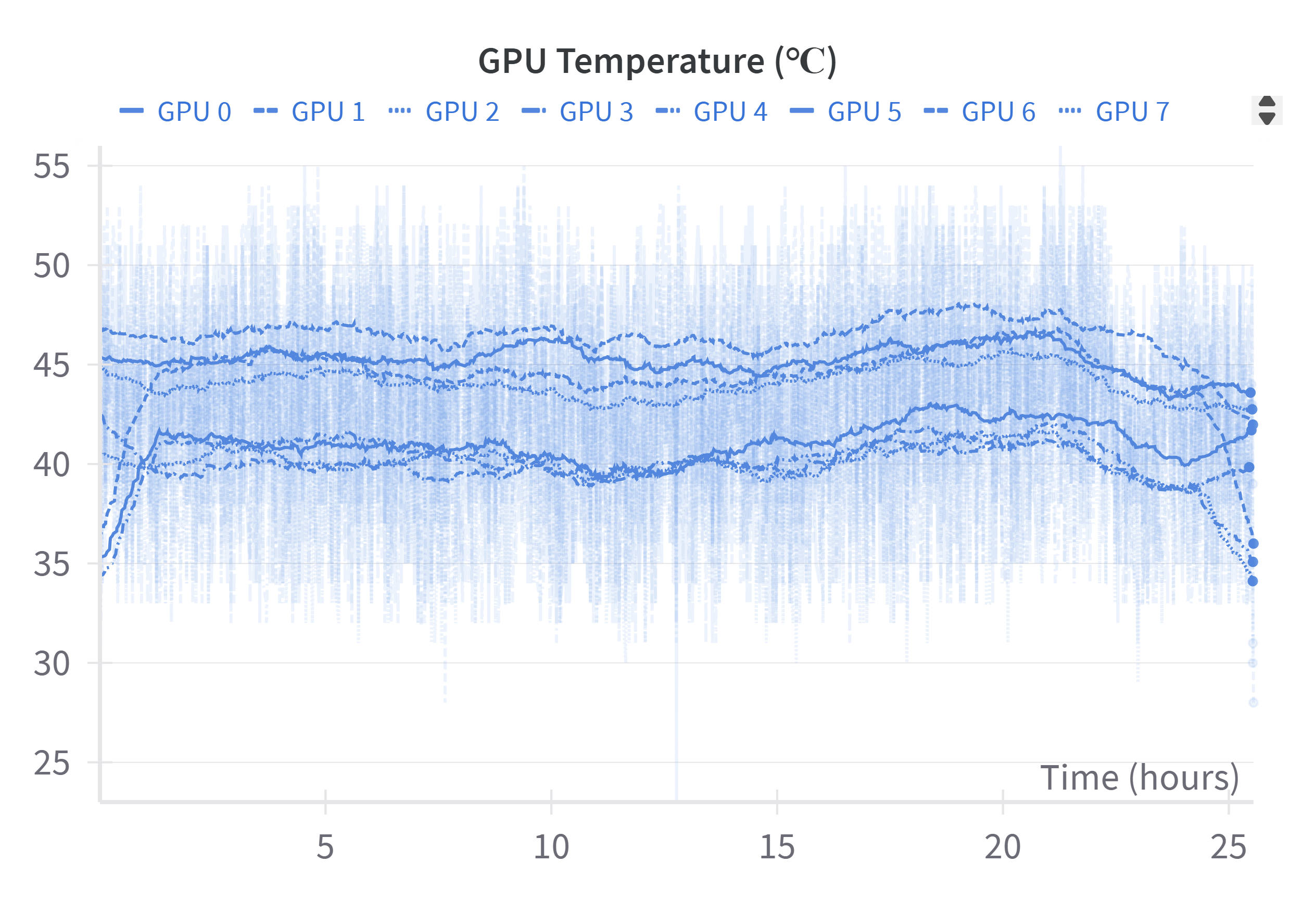}
  }
  \hfill
  \subfloat[\label{fig:vclip_gpu_util_sm}]{%
    \includegraphics[width=0.32\textwidth]{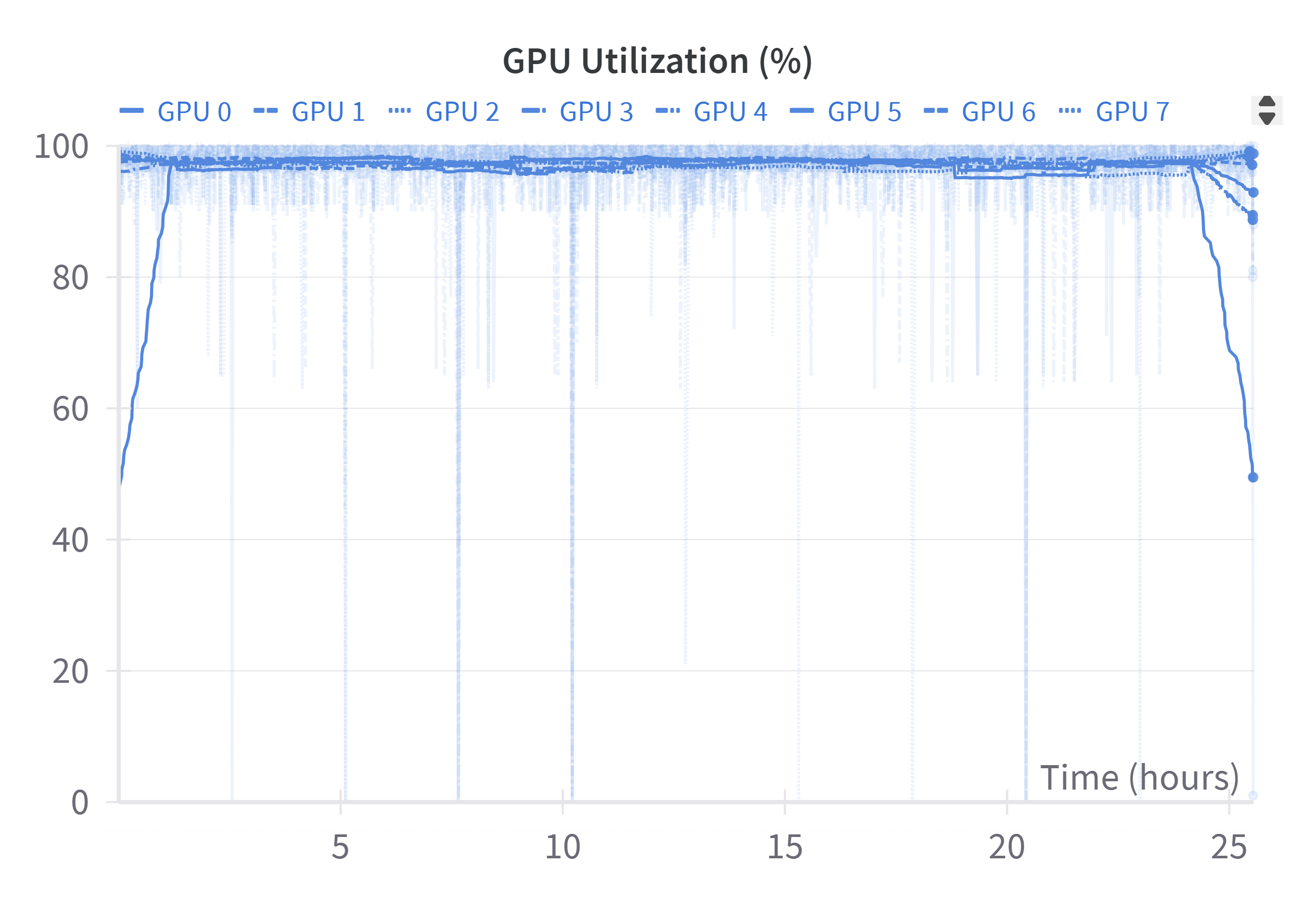}
  }

  \vspace{0.5em}

  \subfloat[\label{fig:vclip_gpu_power_bnl}]{%
    \includegraphics[width=0.32\textwidth]{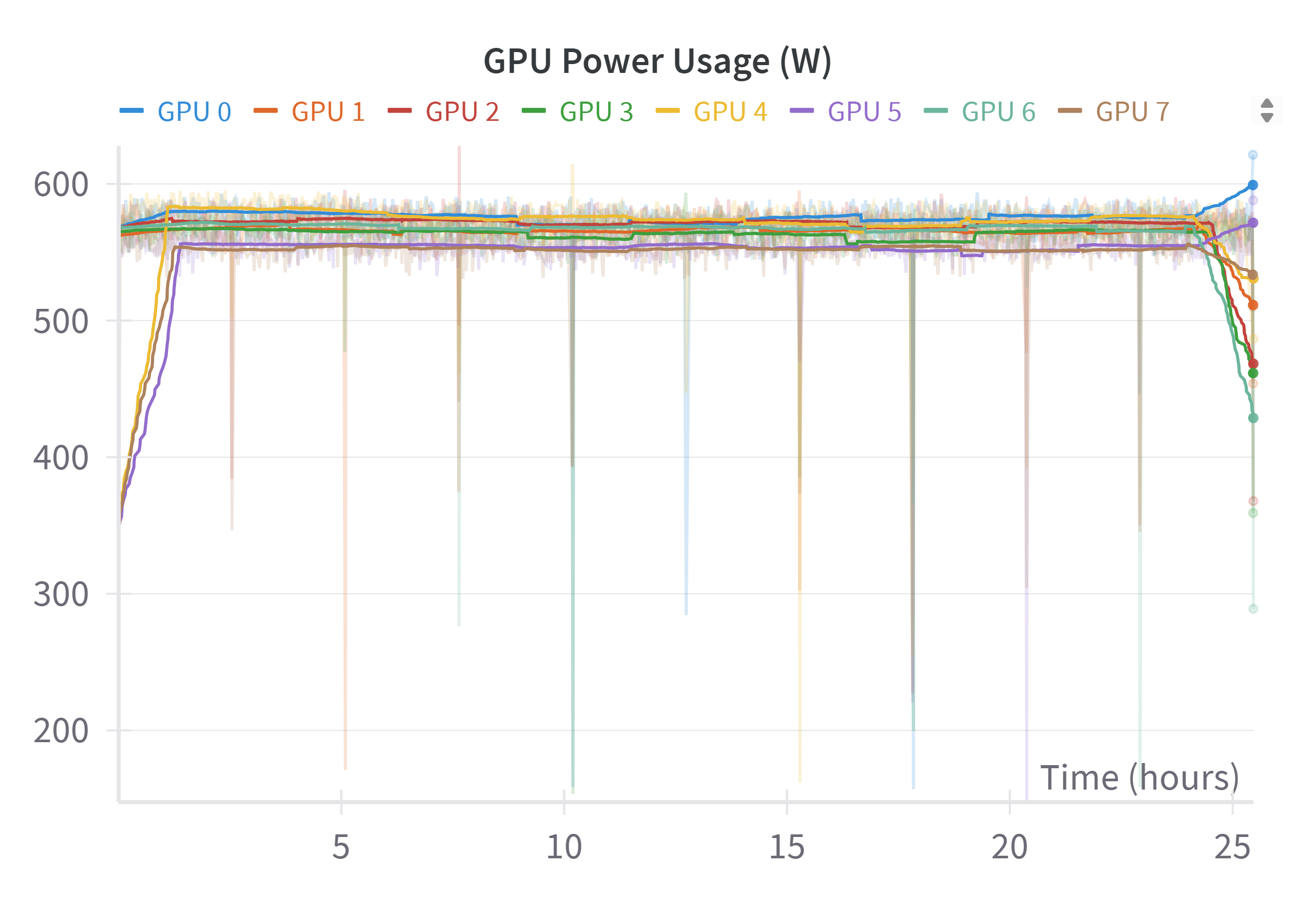}
  }
  \hfill
  \subfloat[\label{fig:vclip_gpu_temp_bnl}]{%
    \includegraphics[width=0.32\textwidth]{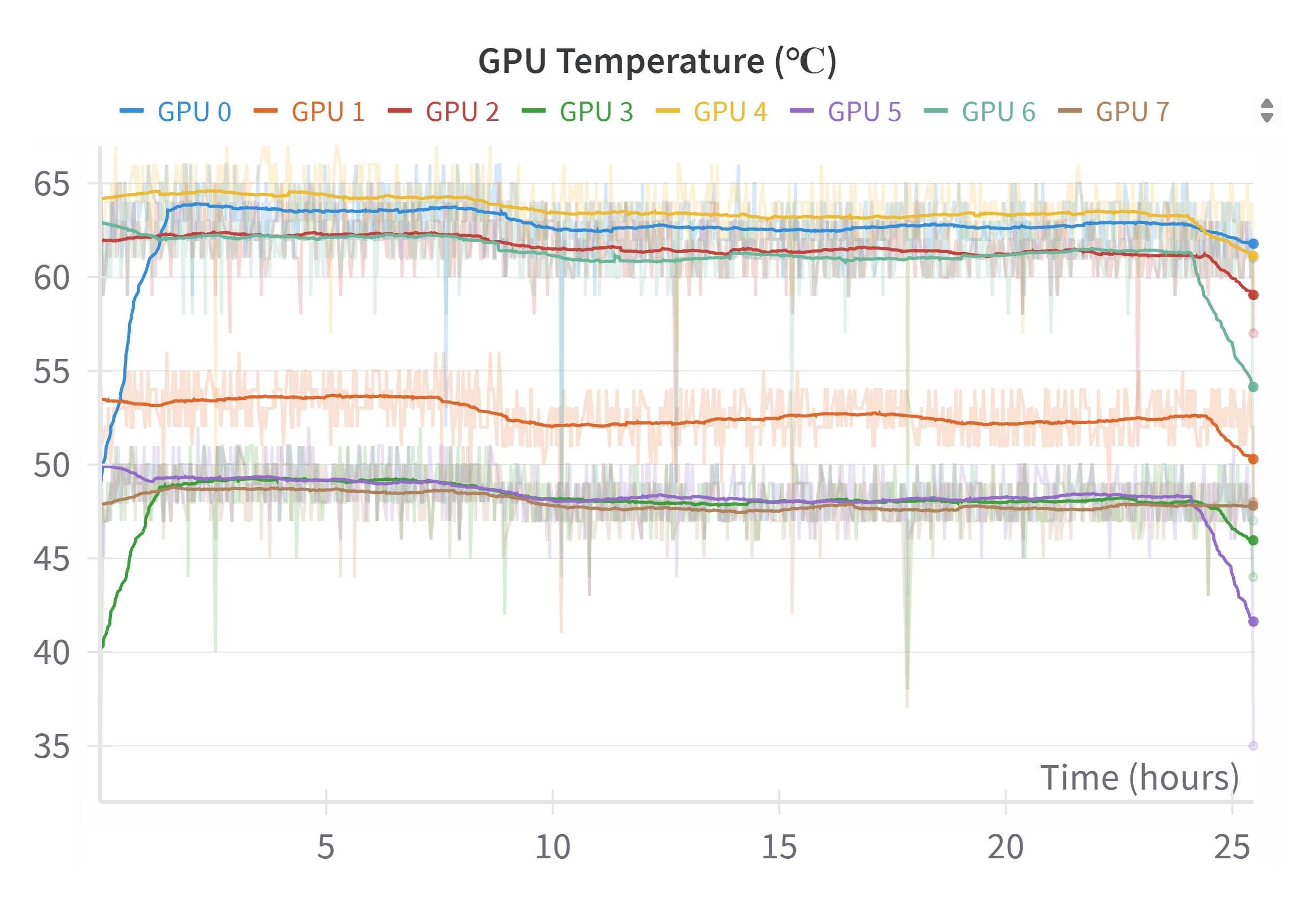}
  }
  \hfill
  \subfloat[\label{fig:vclip_gpu_util_bnl}]{%
    \includegraphics[width=0.32\textwidth]{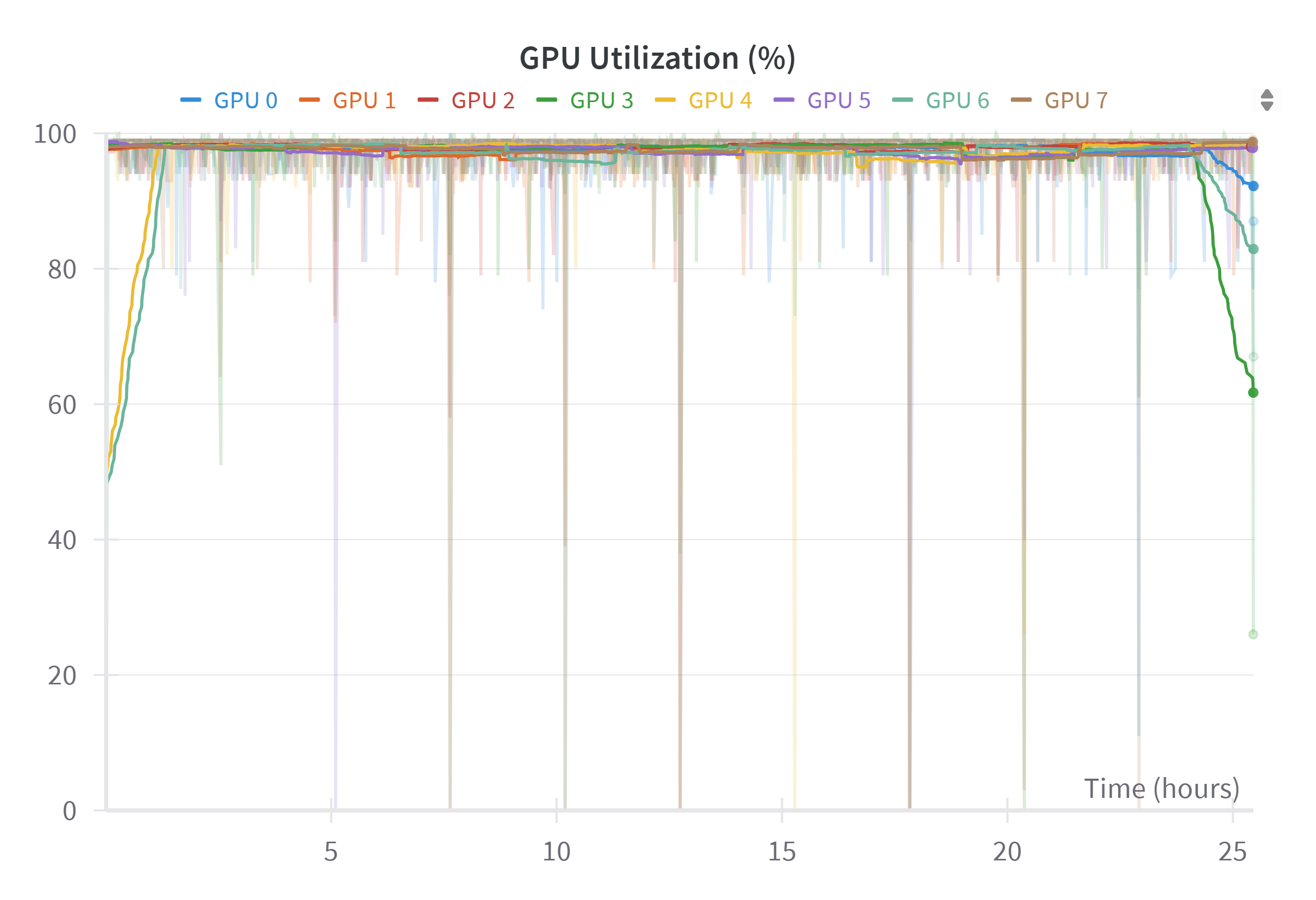}
  }

  \caption{GPU metrics comparison for liquid-cooled (a,b,c) and air-cooled (d,e,f) nodes for Vita-CLIP~\cite{wasim2023vita}.}
  \label{fig:six_figures}
\end{figure*}

\subsection{LLM Results}
We benchmarked five large language models ranging from 7B to 32B parameters on both the liquid-cooled and air-cooled nodes. The liquid-cooled system demonstrated lower power draw and higher performance efficiency across all large language models as illustrated in Table~\ref{tab:llm_training_metrics}. The average GPU power consumption was consistently lower on liquid-cooled node, with values ranging from 3503 W (Mistal-7B-v0.3) to 3742 W (Gemma-2-27B), compared to 3653–3912 W on air-cooled system. This reduction in power consumption corresponded to improved performance-per-watt for liquid-cooled node.\\
\indent In terms of throughput, TFLOPS per GPU were slightly higher on the liquid-cooled system, despite identical training configurations. The largest measurement was observed for the Mistral-Nemo-Base-2407 model, which achieved 35.71 TFLOPs/GPU on liquid-cooled node as compared to 35.52 on air-cooled node. Mistral-7B-v0.3, Llama-3.1-8B, and Gemma-2-27B followed similar trends.\\
\indent Qwen2.5-32B, the largest model in our benchmark, also exhibited high performance for liquid-cooling system. The liquid-cooled node achieved slightly faster training, higher throughput with lower average GPU power consumption (3671 W vs. 3844 W). GPU utilization remained high on both nodes (93.4\% vs. 92.9\%), confirming consistent workload execution as shown in Figure~\ref{fig:six_figures_qwen}. These results reinforce the efficiency benefits of liquid cooling, particularly in better power and thermal performance for large-scale models.\\
\indent Node-level power consumption was also notably lower on the liquid-cooled system with averaging 1000–1200 W less per model. These results confirm that liquid-cooling node delivers significant efficiency gains during real-world LLM finetuning workloads.

\begin{table*}[t]
  \centering
  \caption{Liquid-Cooling Performance Advantage across LLMs and VLMs.}
  \label{tab:performance_advantage}
  \begin{tabular}{lcccc}
    \toprule
    \textbf{System/Model} & \makecell{Duration Reduction \\ (\%)} & \makecell{GPU Power \\ Reduction (W)} & \makecell{Node Power \\ Reduction (W)} & \makecell{Performance (FLOPs) \\ Improvement (\%)} \\
    \midrule
    \multicolumn{5}{c}{\textbf{Large Language Models (LLMs)}} \\
    \midrule
    MISTRAL-7B-V0.3 \cite{jiang2023mistral7b}         & 0.7 & 155 & 1084 & 0.3 \\
    LLAMA-3.1-8B \cite{grattafiori2024llama}          & 0.6 & 148 & 1218 & 1.0 \\
    MISTRAL-NEMO-BASE-2407 \cite{mistralnemo2024}     & 0.4 & 161 & 1244 & 0.5 \\
    GEMMA-2-27B \cite{gemmateam2024gemma2improvingopen} & 0.3 & 170 & 1217 & 0.1 \\
    QWEN2.5-32B \cite{yang2024qwen2}                  & 0.5 & 173 & 1096 & 0.1 \\
    \midrule
    \multicolumn{5}{c}{\textbf{Vision-Language Models (VLMs)}} \\
    \midrule
    X-CLIP \cite{ni2022expanding}                    & 3.0 & 115 & 85   & 0.1 \\
    EVL \cite{lin2022frozen}                         & 5.0 & 31  & 391  & 0.4 \\
    VITA-CLIP \cite{wasim2023vita}                   & 0.3 & 234 & 1508 & 0.1 \\
    \midrule
    \textbf{AVERAGE (LLM \& VLM)}                    & 1.4 & 148 & 980  & 0.3 \\
    \bottomrule
  \end{tabular}
\end{table*}

\subsection{VLM Results}
We benchmarked three VLM models including X-CLIP~\cite{ni2022expanding}, EVL~\cite{lin2022frozen}, and ViTA-CLIP~\cite{wasim2023vita} on two distinct computing nodes: liquid-cooled and air-cooled. Through comprehensive training experiments conducted on both nodes in the same training settings, we observed that the liquid-cooled node consistently exhibited lower power consumption and higher performance efficiency across all three VLM models, as presented in Table~\ref{tab:VLM_results}.\\
\indent As listed in Table~\ref{tab:VLM_results}, the liquid-cooled node consistently outperformed the air-cooled node by exhibiting lower average GPU power for each VLM model used in this work. For instance, X-CLIP consumed 3268 W on the liquid-cooled node and 3383 W on the air-cooled node. Similarly, for EVL and Vita-CLIP models, the liquid-cooled has a lower average GPU power consumption of 2320 W and 4295 W in comparison with the air-cooled which has an average GPU power consumption of 2351 W and 4529 W, respectively.\\
\indent Following the same trend, the liquid-cooled node demonstrates higher GFLOPS/GPU across each VLM model. More specifically, for the X-CLIP mode, the liquid-cooled node achieved 377.13, slightly higher than the air-cooled node’s 376.92 GFLOPS/GPU. Similarly, for the EVL model, the liquid-cooled node obtained 119.87 GFLOPS/GPU in comparison to an air-cooled node which has slightly lower GFLOPS/GPU (119.41). Lastly, Vita-CLIP exhibited 1064.06 on the liquid-cooled node and 1063.81 GFLOPS/GPU on the air-cooled node, thereby, consistently showing the computational efficiency of the liquid-cooled node.\\
\indent From the node-level power consumption results presented in Table~\ref{tab:VLM_results}, it can be observed that the liquid-cooled node consistently demonstrates lower power usage compared to the air-cooled node for each VLM model. For example, in the case of the X-CLIP model, the liquid-cooled node consumed 5890 W average power, which is 85 W less than the 5975 W consumed by the air-cooled node. This difference in node-level power consumption between the two nodes increased for models with longer training durations. Specifically, for the EVL model, the liquid-cooled consumed 4053 W average power, which is 391 W less than the 4444 W consumed by the air-cooled node. Similarly, for the Vita-CLIP model, the liquid-cooled node consumed 5867 W, which is 1508 W lower than the 7375 W consumed by the air-cooled Node.\\
\indent Besides the quantitative analysis, we also performed a visual investigation of the two nodes’ performance in terms of GPU power usage (\%), GPU temperature ($^\circ$C), and GPU utilization (\%), as depicted in~Figure \ref{fig:six_figures}. The first row presents the results for the liquid-cooled node, while the second row illustrates the corresponding results for the air-cooled node. As it can be visually perceived, the liquid-cooled node demonstrates significantly better performance in terms of lower GPU power consumption and reduced GPU temperature while maintaining comparable GPU utilization to the air-cooled node.\\
\indent Overall, these results clearly highlight the advantages of liquid cooling in reducing energy consumption and enhancing computational efficiency for training advanced deep learning models and large vision language models.

\section{Discussion}
\label{sec:discussion}
These results illustrate the complex interaction between thermal design and power efficiency in large-scale AI training infrastructure. Both systems possess identical GPU hardware and performed identical workloads, each cooling strategy produced clear differences in thermal stability, energy consumption, and sustained compute performance. These findings highlight a key insight: as AI models scale in size and complexity, the supporting infrastructure must evolve to maintain efficiency at the system level. Further, the performance of IT hardware is not independent of cooling performance. 

The total required flops of our training workloads fall within the natural run-to-run variability of neural network training when comparing the same workload on different hardware configurations ($\pm 0.05\%$), and are thus appropriate to compare the hardware performance directly. On average, in production workloads, liquid cooling enabled a $.03\%$ increase in per-GPU operations. That throughput performance was most pronounced in the fine-tuning workloads, though also varied considerably workload to workload. This computational speed up resulted in average decrease in training time of $1.34\%$ for each workload on the liquid cooled system. 

The more dramatic increase in throughput during the stress test merits focused discussion. While not perfectly representative of any production workload, stress tests provide valuable bounding and directional insights for the most intensive workloads. The observed $17\%$ increase in computational throughput is most likely the result of the higher cooling performance enabling a higher computational clockspeed. Nvidia H100s operate with dynamic-voltage-frequency-scaling, where the GPU frequency is automatically adjusted based on temperature and workload characteristics. The lower average GPU temperature during the stress test enabled faster computation without jeopardizing hardware health. In production workloads, this would enable improved energy performance through reduction in computational time in addition to any hardware or infrastructure power difference. 

The liquid-cooled system consistently maintained lower and more stable GPU temperatures under load, directly contributing to higher compute throughput and reduced power draw.  These improvements cannot be attributed to GPU variance alone; rather, they reflect the systemic benefit of shifting thermal load from less efficient air-cooling mechanisms to direct-to-chip liquid cooling. This reinforces the idea that cooling is no longer a passive operational concern, but an active design variable that influences the performance-per-watt equation in high-density compute environments.

While the observed improvements in computational throughput are meaningful, especially given the scale of industrial clusters and the workloads they service, the most dramatic difference in node-level performance was in power demand. On average, the liquid cooled node drew ~5.2 kW during production workloads, as compared to ~6.2 kW by the air cooled node. There is considerable variability by computational intensity: the lowest GPU-utilization workloads, EVL and X-CLIP, differed only modestly in their power demand (100-400 watts). The language model fine tuning workloads averaged 1.2 kW lower instantaneous power demand on the liquid cooled node, with the highest-utilization workload of Vita-CLIP demanding a full 1.5 kW less on the liquid cooled node as shown in Table~\ref{tab:performance_advantage}. These values at high utilization are consistent with the difference in node power demand observed during the stress test. 

The power savings from liquid cooling become increasingly significant as computational load and power draw increase, with the most dramatic differences observed during high-utilization workloads. This relationship stems primarily from the fundamental thermal efficiency differences between air and liquid cooling systems. Air-cooled nodes contain embedded cooling fans, which are supplied power at the node-level. These fans remain minimally active during low-utilization tasks when thermal management demands are modest, resulting in relatively small power differentials between cooling approaches (only 100-400 watts for low-GPU-utilization workloads like EVL and X-CLIP). However, as computational intensity rises, the limitations of air cooling become increasingly apparent. During high-utilization workloads like language model fine-tuning and Vita-CLIP, the air-cooled node's fans must work substantially harder, consuming more power themselves while being less effective at heat removal than direct-to-chip liquid cooling, creating a compounding efficiency gap. This drives the observed 1.2-1.5 kW power reduction in liquid-cooled systems during intensive tasks. 

Note that these embedded fans skew PUE as typically measured. Because they're supplied power at the node-level, their power demand is considered IT load for facility metering and monitoring. This means the cooling performance improvements from liquid cooling offer a dual benefit: they not only reduce absolute power consumption but also effectively transfer cooling workload from less efficient embedded fans (counted as IT load) to more efficient centralized chilled water systems (counted as facility overhead). This transfer of thermal management responsibility produces a more accurate representation of true infrastructure efficiency while simultaneously improving overall system performance. The energy-intensive, distributed cooling load previously hidden within IT power metrics becomes managed by purpose-built central cooling infrastructure with superior coefficient of performance. The net result is both reduced total energy consumption and a more transparent assessment of facility cooling efficiency that better reflects the actual relationship between productive computation and supporting infrastructure.

Another key observation is the sensitivity of total energy consumption to workload configuration. While batch size and finetuning method were held consistent within workloads to adequately compare the hardware, architectural differences among LLMs and VLMs influenced both power demand and compute efficiency. Additionally, the fine-tuning task may present differential power characteristics as compared to pre-training, introducing a potential confounding factor into these observed architectural effects. This highlights the need for workload-aware optimization strategies that consider not just model accuracy, but also energy footprint, a perspective increasingly relevant as AI scales to gigawatt clusters.

However, these findings must be interpreted within the context of the specific configurations tested. The systems differed not only in cooling but also in CPU architecture and platform integration, introducing confounding variables that may affect node-level power behavior. Although GPU power is the majority of node power demand, other components contribute meaningful variability. Despite this, the convergence of our stress test results near the manufacturer-rated maximums suggests that our benchmarks capture representative upper bounds for 8$\times$ H100 deployments.

Moreover, while this study focused on single-node analysis, real-world training frequently spans multiple nodes or even distributed datacenters. Multi-node scaling introduces non-trivial communication overheads, synchronization latency, and network power draw, all of which can dilute the gains observed in isolated systems. Contrastingly, the scale economies of cooling systems and coolant delivery infrastructure may unlock improved infrastructure effeciency.  Future studies should investigate multi-node scenarios, measuring the energy impact of parallelization strategies and cluster-level orchestration, as well as system wide energy performance.

Finally, the absence of fine-grained sub-metering remains a limitation in understanding component-level power distribution. While our use of IPMI and software-based profilers enabled high-level insights, detailed sub-component power attribution, down to memory banks, PCIe links, or fan curves, would offer a more precise picture of system inefficiencies and optimization opportunities. As hardware manufacturers move toward integrated power telemetry, we expect future research to enable more complex component-level energy optimizations.

In summary, this work provides strong empirical evidence that liquid cooling yields significant gains in performance-per-watt and thermal consistency for modern AI workloads. As models continue to scale and datacenters approach thermal and electrical limits, rethinking cooling as a primary architectural variable will be essential. The future of sustainable AI depends not only on model and algorithmic innovations, but also on the physical infrastructure that enables them.

\section{Conclusion}
\label{sec:conclusion}
This study presents a comparative analysis of air-cooled and liquid-cooled GPU systems for training large language and vision-language models. By benchmarking across identical GPU hardware configurations each equipped with 8× NVIDIA H100 GPUs, we found that liquid-cooled systems consistently outperformed their air-cooled counterparts in thermal stability, power efficiency, and throughput.

Liquid-cooled system maintained GPU temperatures within a narrower and lower range (41–50°C), while air-cooled system exhibited greater variability and reached peak temperatures above 70°C. This thermal advantage translated into approximately 17\% higher TFLOPS per GPU under GPU Burn test, and similar trend was observed during model training. Additionally, node-level power consumption was significantly lower in liquid-cooled systems, indicating improved performance-per-watt across all workloads.

These results highlight the critical role of thermal design in optimizing AI infrastructure at scale. As large model training continues to grow in complexity and intensity, the adoption of liquid cooling offers a promising path forward for sustainable and energy-efficient data center operations.

\section*{Acknowledgment}

During the preparation of this work, the author(s) used ChatGPT-4o to improve language clarity and typesetting. After using this tool, the author(s) reviewed and edited the content as necessary and took full responsibility for the final version of the publication.

\bibliographystyle{IEEEtran}
\bibliography{References}

\begin{thebibliography}{10}
\providecommand{\url}[1]{#1}
\csname url@samestyle\endcsname
\providecommand{\newblock}{\relax}
\providecommand{\bibinfo}[2]{#2}
\providecommand{\BIBentrySTDinterwordspacing}{\spaceskip=0pt\relax}
\providecommand{\BIBentryALTinterwordstretchfactor}{4}
\providecommand{\BIBentryALTinterwordspacing}{\spaceskip=\fontdimen2\font plus
\BIBentryALTinterwordstretchfactor\fontdimen3\font minus \fontdimen4\font\relax}
\providecommand{\BIBforeignlanguage}[2]{{%
\expandafter\ifx\csname l@#1\endcsname\relax
\typeout{** WARNING: IEEEtran.bst: No hyphenation pattern has been}%
\typeout{** loaded for the language `#1'. Using the pattern for}%
\typeout{** the default language instead.}%
\else
\language=\csname l@#1\endcsname
\fi
#2}}
\providecommand{\BIBdecl}{\relax}
\BIBdecl

\bibitem{kaplan2020scalinglawsneurallanguage1}
\BIBentryALTinterwordspacing
J.~Kaplan, S.~McCandlish, T.~Henighan, T.~B. Brown, B.~Chess, R.~Child, S.~Gray, A.~Radford, J.~Wu, and D.~Amodei, ``Scaling laws for neural language models,'' 2020. [Online]. Available: \url{https://arxiv.org/abs/2001.08361}
\BIBentrySTDinterwordspacing

\bibitem{rattner2024aiboom2}
\BIBentryALTinterwordspacing
N.~Rattner and T.~Dotan, ``The ai spending boom in charts,'' \emph{The Wall Street Journal}, September 2024, tech Section. [Online]. Available: \url{https://www.wsj.com/tech/ai/artificial-intelligence-investing-charts-7b8e1a97}
\BIBentrySTDinterwordspacing

\bibitem{schmidt2024gpus3}
B.~Schmidt and A.~Hildebrandt, ``From gpus to ai and quantum: three waves of acceleration in bioinformatics,'' \emph{Drug Discovery Today}, p. 103990, 2024.

\bibitem{nave2021artificial4}
O.~Nave and M.~Elbaz, ``Artificial immune system features added to breast cancer clinical data for machine learning (ml) applications,'' \emph{Biosystems}, vol. 202, p. 104341, 2021.

\bibitem{kaack2022aligning5}
\BIBentryALTinterwordspacing
L.~H. Kaack, P.~L. Donti, E.~Strubell, G.~Kamiya, F.~Creutzig, and D.~Rolnick, ``Aligning artificial intelligence with climate change mitigation,'' \emph{Nature Climate Change}, vol.~12, no.~6, pp. 518--527, 2022. [Online]. Available: \url{https://doi.org/10.1038/s41558-022-01377-7}
\BIBentrySTDinterwordspacing

\bibitem{narayanan2024investigation}
A.~Narayanan, Q.~Wang, S.~Ozguc, and R.~W. Bonner~III, ``Investigation of server level direct-to-chip two phase cooling solution for high power gpus,'' in \emph{International Electronic Packaging Technical Conference and Exhibition}, vol. 88469.\hskip 1em plus 0.5em minus 0.4em\relax American Society of Mechanical Engineers, 2024, p. V001T02A009.

\bibitem{ohenhen2024sustainable}
P.~E. Ohenhen, O.~Chidolue, A.~A. Umoh, B.~Ngozichukwu, A.~V. Fafure, V.~I. Ilojianya, and K.~I. Ibekwe, ``Sustainable cooling solutions for electronics: A comprehensive review: Investigating the latest techniques and materials, their effectiveness in mechanical applications, and associated environmental benefits,'' \emph{World Journal of Advanced Research and Reviews}, vol.~21, no.~1, pp. 957--972, 2024.

\bibitem{zhou2024immersion}
K.~Zhou, X.~Yu, B.~Xie, H.~Xie, and W.~Fu, ``Immersion cooling technology development status of data center,'' \emph{Science and Technology for Energy Transition}, vol.~79, p.~41, 2024.

\bibitem{naduvilakath2024numerical}
F.~Naduvilakath-Mohammed, M.~Lebon, and A.~Robinson, ``Numerical modelling of a hybrid vapor compression refrigeration assisted closed loop liquid cooling system for high-performance computing systems,'' in \emph{Journal of Physics: Conference Series}, vol. 2766, no.~1.\hskip 1em plus 0.5em minus 0.4em\relax IOP Publishing, 2024, p. 012078.

\bibitem{wang2024research}
Y.~Wang, ``Research on key technology and system application of energy management coupled with liquid cooling and intelligent control algorithm,'' in \emph{E3S Web of Conferences}, vol. 520.\hskip 1em plus 0.5em minus 0.4em\relax EDP Sciences, 2024, p. 04020.

\bibitem{shao2020evaluation}
S.~Shao, T.~Gao, H.~Yang, J.~Zhao, and J.~Zhang, ``Evaluation of single phase immersion cooling system for high performance server chassis using dielectric coolants,'' in \emph{International Electronic Packaging Technical Conference and Exhibition}, vol. 84041.\hskip 1em plus 0.5em minus 0.4em\relax American Society of Mechanical Engineers, 2020, p. V001T08A010.

\bibitem{jiang2023mistral7b}
\BIBentryALTinterwordspacing
A.~Q. Jiang, A.~Sablayrolles, A.~Mensch, C.~Bamford, D.~S. Chaplot, D.~de~las Casas, F.~Bressand, G.~Lengyel, G.~Lample, L.~Saulnier, L.~R. Lavaud, M.-A. Lachaux, P.~Stock, T.~L. Scao, T.~Lavril, T.~Wang, T.~Lacroix, and W.~E. Sayed, ``Mistral 7b,'' 2023. [Online]. Available: \url{https://arxiv.org/abs/2310.06825}
\BIBentrySTDinterwordspacing

\bibitem{grattafiori2024llama}
A.~Grattafiori, A.~Dubey, A.~Jauhri, A.~Pandey, A.~Kadian, A.~Al-Dahle, A.~Letman, A.~Mathur, A.~Schelten, A.~Vaughan \emph{et~al.}, ``The llama 3 herd of models,'' \emph{arXiv preprint arXiv:2407.21783}, 2024.

\bibitem{mistralnemo2024}
\BIBentryALTinterwordspacing
{Mistral AI Team}, ``Mistral nemo: Our new best small model,'' \url{https://mistral.ai/news/mistral-nemo}, July 2024, accessed March 2025. [Online]. Available: \url{https://mistral.ai/news/mistral-nemo}
\BIBentrySTDinterwordspacing

\bibitem{gemmateam2024gemma2improvingopen}
\BIBentryALTinterwordspacing
G.~Team, M.~Riviere, S.~Pathak, P.~G. Sessa, C.~Hardin, S.~Bhupatiraju \emph{et~al.}, ``Gemma 2: Improving open language models at a practical size,'' 2024. [Online]. Available: \url{https://arxiv.org/abs/2408.00118}
\BIBentrySTDinterwordspacing

\bibitem{yang2024qwen2}
A.~Yang, B.~Yang, B.~Zhang, B.~Hui, B.~Zheng, B.~Yu, C.~Li, D.~Liu, F.~Huang, H.~Wei \emph{et~al.}, ``Qwen2. 5 technical report,'' \emph{arXiv preprint arXiv:2412.15115}, 2024.

\bibitem{ni2022expanding}
B.~Ni, H.~Peng, M.~Chen, S.~Zhang, G.~Meng, J.~Fu, S.~Xiang, and H.~Ling, ``Expanding language-image pretrained models for general video recognition,'' in \emph{European conference on computer vision}.\hskip 1em plus 0.5em minus 0.4em\relax Springer, 2022, pp. 1--18.

\bibitem{lin2022frozen}
Z.~Lin, S.~Geng, R.~Zhang, P.~Gao, G.~De~Melo, X.~Wang, J.~Dai, Y.~Qiao, and H.~Li, ``Frozen clip models are efficient video learners,'' in \emph{European Conference on Computer Vision}.\hskip 1em plus 0.5em minus 0.4em\relax Springer, 2022, pp. 388--404.

\bibitem{wasim2023vita}
S.~T. Wasim, M.~Naseer, S.~Khan, F.~S. Khan, and M.~Shah, ``Vita-clip: Video and text adaptive clip via multimodal prompting,'' in \emph{Proceedings of the IEEE/CVF Conference on Computer Vision and Pattern Recognition}, 2023, pp. 23\,034--23\,044.

\bibitem{kim2024data}
J.~H. Kim, D.~U. Shin, and H.~Kim, ``Data center energy evaluation tool development and analysis of power usage effectiveness with different economizer types in various climate zones,'' \emph{Buildings}, vol.~14, no.~1, p. 299, 2024.

\bibitem{tang2023experimental}
Y.~Tang, X.~Zhang, and Z.~Liu, ``Experimental study on the thermal performance of flat loop heat pipe applied in data center cooling,'' \emph{Energies}, vol.~16, no.~12, p. 4677, 2023.

\bibitem{haghshenas2023enough}
K.~Haghshenas, B.~Setz, Y.~Blosch, and M.~Aiello, ``Enough hot air: the role of immersion cooling,'' \emph{Energy Informatics}, vol.~6, no.~1, p.~14, 2023.

\bibitem{icae2021freecooling}
M.~Borkowski and A.~Piłat, ``Saving energy and efficiency increase by enabling free-cooling mode for cooling system in data center,'' in \emph{Proceedings of the 13th International Conference on Applied Energy (ICAE2021)}, ser. Energy Proceedings, vol.~23.\hskip 1em plus 0.5em minus 0.4em\relax ICAE, November 2021, iCAE International Conference on Applied Energy, Nov. 29–Dec. 5, 2021, Thailand/Virtual. Paper ID: 240.

\bibitem{gupta2021energy}
R.~Gupta, S.~Asgari, H.~Moazamigoodarzi, D.~G. Down, and I.~K. Puri, ``Energy, exergy and computing efficiency based data center workload and cooling management,'' \emph{Applied Energy}, vol. 299, p. 117050, 2021.

\bibitem{wang2022toward}
R.~Wang, X.~Zhang, X.~Zhou, Y.~Wen, and R.~Tan, ``Toward physics-guided safe deep reinforcement learning for green data center cooling control,'' in \emph{2022 ACM/IEEE 13th International Conference on Cyber-Physical Systems (ICCPS)}.\hskip 1em plus 0.5em minus 0.4em\relax IEEE, 2022, pp. 159--169.

\bibitem{athavale2021}
J.~Athavale, M.~Yoda, and Y.~Joshi, ``Genetic algorithm based cooling energy optimization of data centers,'' \emph{International Journal of Numerical Methods for Heat \& Fluid Flow}, vol.~31, pp. 3148--3168, 2021.

\bibitem{ebirim2024optimizing}
W.~Ebirim, F.~Usman, K.~Olu-Lawal, N.~Ninduwesuor-Ehiobu, E.~Ani, and D.~Montero, ``Optimizing energy efficiency in data center cooling towers through predictive maintenance and project management,'' \emph{World Journal of Advanced Research and Reviews}, vol.~21, no.~2, pp. 1782--1790, 2024.

\bibitem{zhao2023simulation}
T.~Zhao, R.~Sun, X.~Hou, J.~Huang, W.~Geng, and J.~Jiang, ``Simulation study of influencing factors of immersion phase-change cooling technology for data center servers,'' \emph{Energies}, vol.~16, no.~12, p. 4640, 2023.

\bibitem{jia2024simulation}
D.~Jia, W.~He, C.~Xu, and T.~Guo, ``Simulation study of modular independent cooling systems for servers,'' in \emph{Journal of Physics: Conference Series}, vol. 2835, no.~1.\hskip 1em plus 0.5em minus 0.4em\relax IOP Publishing, 2024, p. 012066.

\bibitem{ipmitool}
{ipmitool contributors}, ``ipmitool - utility for ipmi-enabled systems,'' \url{https://github.com/ipmitool/ipmitool}, 2025, accessed: April 11, 2025.

\bibitem{wandb}
\BIBentryALTinterwordspacing
L.~Biewald, ``Experiment tracking with weights and biases,'' 2020, software available from wandb.com. [Online]. Available: \url{https://www.wandb.com/}
\BIBentrySTDinterwordspacing

\bibitem{benoit_courty_2024_11171501}
\BIBentryALTinterwordspacing
B.~Courty, V.~Schmidt, S.~Luccioni, Goyal-Kamal, MarionCoutarel, B.~Feld, J.~Lecourt, LiamConnell, A.~Saboni, Inimaz, supatomic, M.~Léval, L.~Blanche, A.~Cruveiller, ouminasara, F.~Zhao, A.~Joshi, A.~Bogroff, H.~de~Lavoreille, N.~Laskaris, E.~Abati, D.~Blank, Z.~Wang, A.~Catovic, M.~Alencon, M.~Stęchły, C.~Bauer, L.~O.~N. de~Araújo, JPW, and MinervaBooks, ``mlco2/codecarbon: v2.4.1,'' May 2024. [Online]. Available: \url{https://doi.org/10.5281/zenodo.11171501}
\BIBentrySTDinterwordspacing

\bibitem{timonen2024gpuburn}
\BIBentryALTinterwordspacing
V.~Timonen, ``wilicc/gpu-burn,'' \url{https://github.com/wilicc/gpu-burn}, October 2024, original-date: 2017-11-25. [Online]. Available: \url{https://github.com/wilicc/gpu-burn}
\BIBentrySTDinterwordspacing

\bibitem{alpaca}
R.~Taori, I.~Gulrajani, T.~Zhang, Y.~Dubois, X.~Li, C.~Guestrin, P.~Liang, and T.~B. Hashimoto, ``Stanford alpaca: An instruction-following llama model,'' \url{https://github.com/tatsu-lab/stanford_alpaca}, 2023.

\bibitem{soomro2012ucf101}
K.~Soomro, A.~R. Zamir, and M.~Shah, ``Ucf101: A dataset of 101 human actions classes from videos in the wild,'' \emph{arXiv preprint arXiv:1212.0402}, 2012.

\end{thebibliography}
\vskip -2\baselineskip plus -1fil

\begin{IEEEbiography}[{\includegraphics[width=1.05in,height=1.3in,clip]{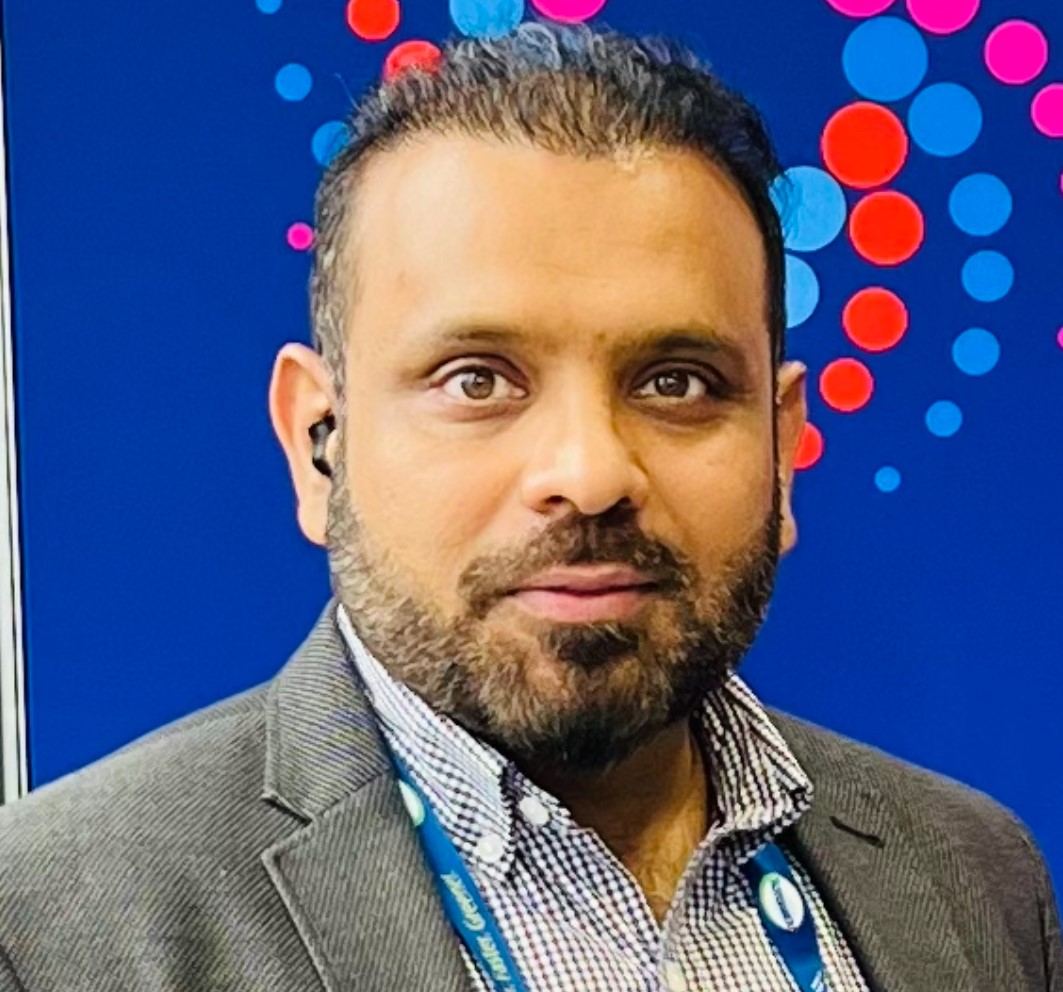}}]{Imran Latif} (Member, IEEE) received the master's degree in mechanical engineering from The City College of New York. He is currently Vice President of Global Technology and Innovation, Data Centers, at Johnson Controls. Before joining Johnson Controls, he was an Engineering Executive and the Chief Operations Officer with the Brookhaven National Laboratory, where he led the Infrastructure Operations of High Performance Computing Centers, supported the global cutting-edge research collaborations on physics, life sciences, quantum, AI, and HPC. He has over 20 years of leadership experience delivering large scale engineering, construction, and sustainability projects. He leads worldwide sustainability research with leading academia and tech companies. His research work is focused on solving the real-life energy issues in the mission critical facilities and developing the AI and ML-based technologies to automate the sustainability processes. He is a subject matter Expert in direct to chip and immersion cooling in data centers.\end{IEEEbiography}

\begin{IEEEbiography}[{\includegraphics[width=1.05in,height=1.3in,clip, keepaspectratio]{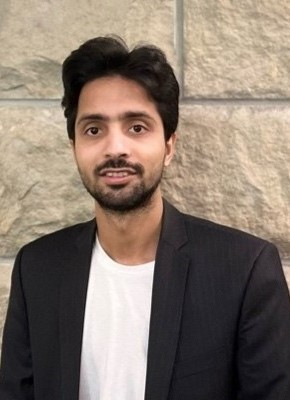}}]{Muhammad Ali Shafique}
received the Bachelor of Science and master’s degrees from the University of Engineering and Technology, Lahore, in 2015 and 2017, respectively. He is currently pursuing the Ph.D. degree in Electrical and Computer Engineering at Kansas State University. He also holds the position of a Research Assistant with the ISCAAS Laboratory. His research interest includes design of efficient LLMs in resource-constrained systems. He is committed to conducting research in this area with the aim of advancing the current knowledge and understanding of these fields.\end{IEEEbiography}

\begin{IEEEbiography}[{\includegraphics[width=1.05in,height=1.3in,clip, keepaspectratio]{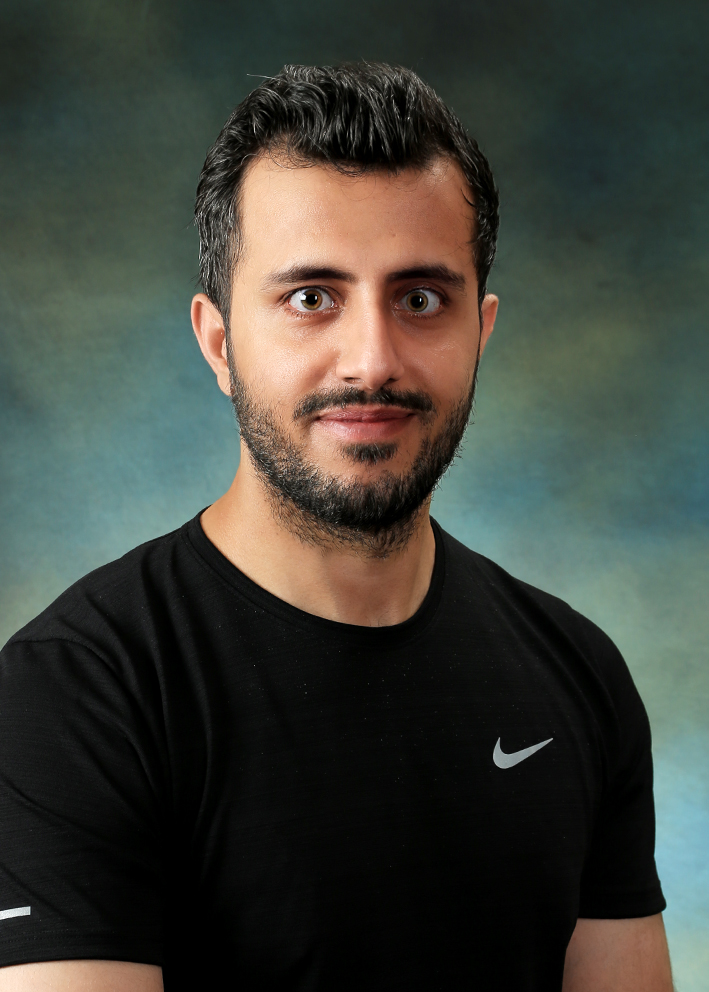}}]{Hayat Ullah}
received his Bachelor's degree in Computer Science from Islamia College University, Peshawar, Pakistan, in 2018, and his Master's degree in Computer Science from Sejong University, Seoul, Republic of Korea, in 2021. He is currently pursuing his Ph.D. in Computer Science at Florida Atlantic University, Boca Raton, FL, USA. He is also a Research Assistant with the Intelligent Systems, Computer Architecture, Analytics, and Security (ISCAAS) Laboratory, Florida Atlantic University. He is exclusively working on multi-model human actions modeling and activity recognition. He has published several articles in well-reputed journals, including IEEE Internet of Things Journal and IEEE Transactions on Image Processing. His research interests include human action recognition, temporal action localization, knowledge distillation, adversarial robustness, vision-language models for video analytics, distributed training, and HPC nodes performance benchmarking using LLMs and VLMs.\end{IEEEbiography}

\begin{IEEEbiography}[{\includegraphics[width=1.05in,height=1.3in,clip, keepaspectratio]{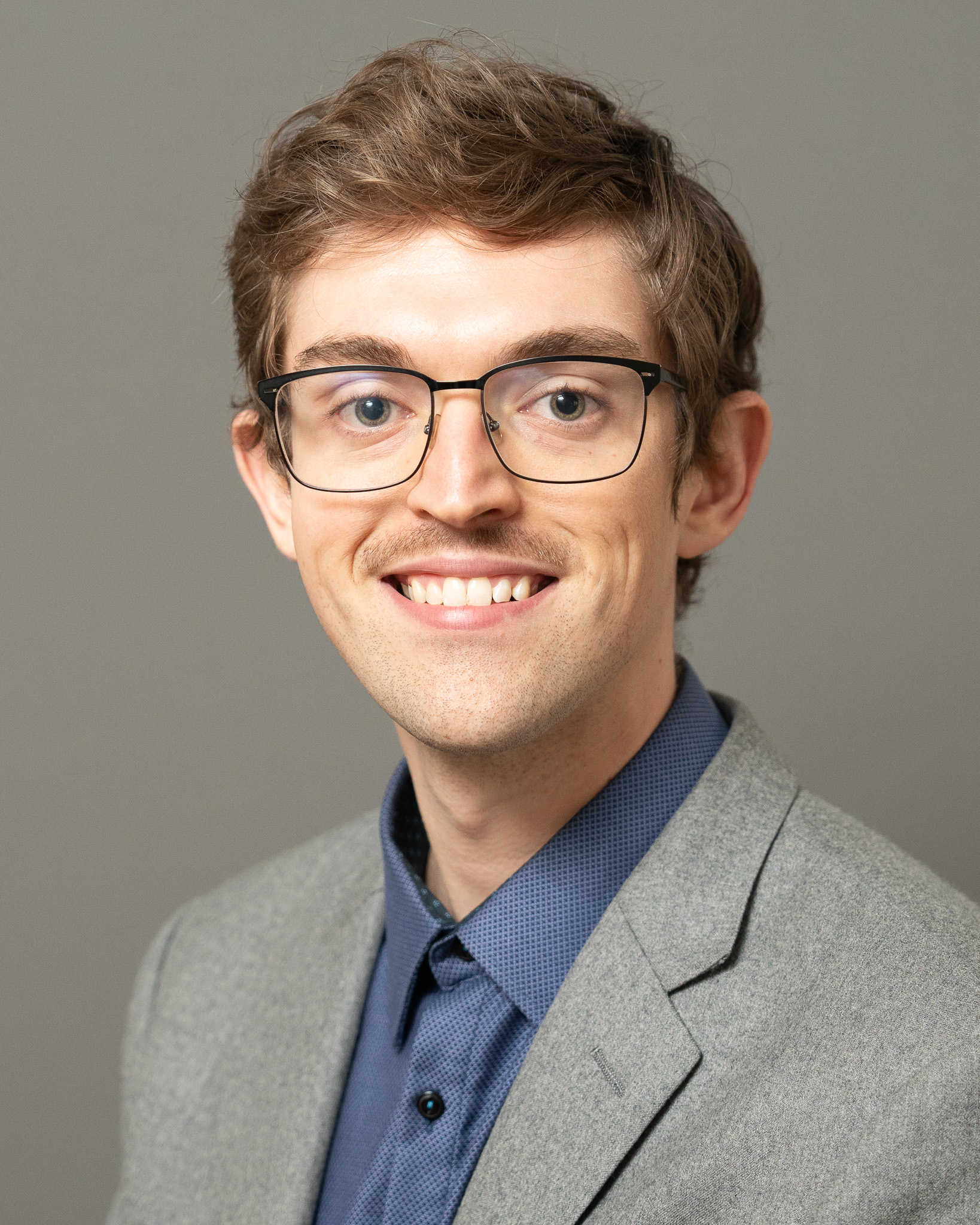}}]{ALEX C. NEWKIRK}
received the bachelor’s degree in physics from Carleton College and the Ph.D. degree in engineering and public policy researching the criticality of semiconductors from Carnegie Mellon University. He is an Energy Technology Researcher with the Lawrence Berkeley National Laboratory and a member of the Center of Expertise for Energy Efficiency in Data Centers. He has led or contributed to research on organizational barriers to energy efficiency and AI related computational demand.\end{IEEEbiography}

\begin{IEEEbiography}[{\includegraphics[width=1.05in,height=1.3in,clip,keepaspectratio]{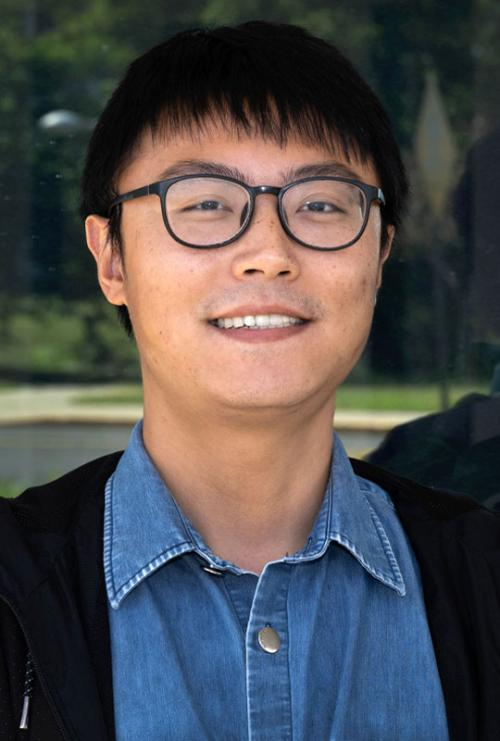}}]{Xi YU}
(Member, IEEE) received the master’s and Ph.D. degrees in electrical and computer engineering from the University of Florida, in 2019 and 2022, respectively. He is currently a Postdoctoral Research Associate with the Computing and Data Sciences Directorate, Brookhaven National Laboratory. His research interests include domain generalization, robust machine learning, information theory, and scientific imaging applications.\end{IEEEbiography}

\begin{IEEEbiography}[{\includegraphics[width=1.05in,
height=1.3in,clip, keepaspectratio]{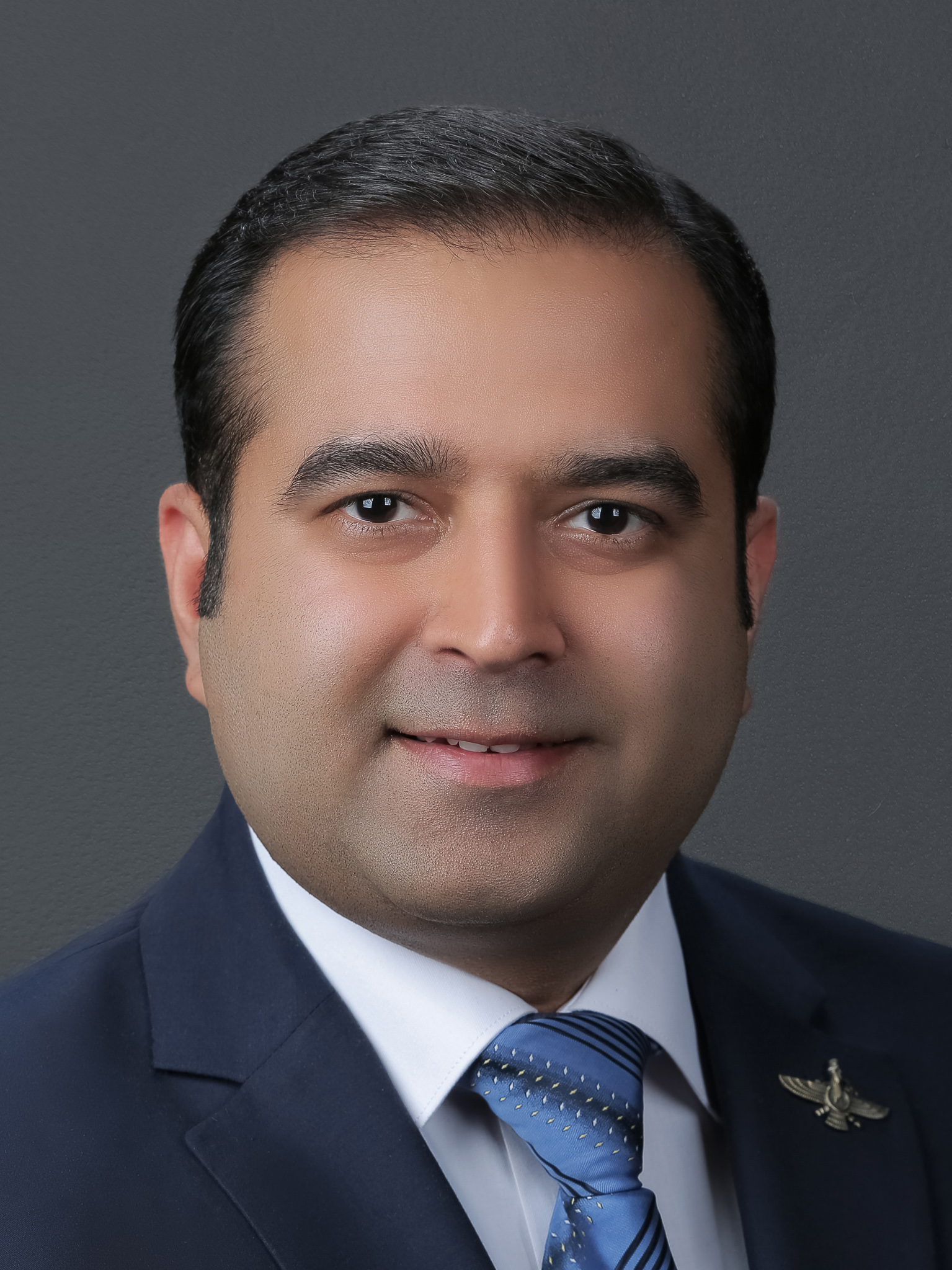}}]{Arslan Munir} (M'09,
SM'17) received his M.A.Sc. degree in electrical and computer engineering (ECE) from the University of British Columbia, Vancouver, Canada, in 2007, and his Ph.D. degree in ECE from the University of Florida, Gainesville, FL, USA, in 2012. From 2007 to 2008, he worked as a Software Development Engineer at the Embedded Systems Division, Mentor Graphics Corporation. He was a Postdoctoral Research Associate with the ECE Department at Rice University, Houston, TX, USA, from May 2012 to June 2014. He is currently an Associate Professor in the Department of Electrical Engineering and Computer Science at Florida Atlantic University. He is also an Adjunct Associate Professor of Computer Science and the University Outstanding Scholar at Kansas State University. 

Munir's current research interests include embedded and cyber-physical systems, secure and trustworthy systems, parallel computing, artificial intelligence, and computer vision. He has received many academic awards, including the Doctoral Fellowship from the Natural Sciences and Engineering Research Council (NSERC) of Canada. He earned gold medals for best performance in electrical engineering and gold medals and academic roll of honor for securing rank one in pre-engineering provincial examinations (out of approximately 300,000
candidates).
\end{IEEEbiography}
\end{document}